\documentclass[12pt]{article}
\usepackage{amsfonts,amsthm,amsmath,amssymb,upgreek}
\usepackage{hyperref}
\usepackage{cite}
\usepackage[paper=letterpaper,margin=1in]{geometry}
\usepackage{graphicx}
\usepackage{units}
\usepackage{dsdshorthand}
\usepackage{color}

\def\hat{\widehat}
\renewcommand\bar{\overline}
\renewcommand{\be}{\begin{equation}}
\newcommand{\ee}{\end{equation}}

\def\b{\overline}

\newcommand\vol{\mathrm{vol}}

\numberwithin{equation}{section}

\begin{document}
\thispagestyle{empty}

\begin{center}

{\bf {\LARGE A spacetime derivation of the Lorentzian \\ \vspace{10pt} OPE inversion formula}\\

\vspace{1cm}}

  \begin{center}

 {\bf David Simmons-Duffin,$^{a,b}$ Douglas Stanford,$^b$  and Edward Witten$^b$}\\
  \bigskip \rm
  
\bigskip
 $^a$Walter Burke Institute for Theoretical Physics, Caltech, Pasadena, CA 91125  \\
   $^b$Institute for Advanced Study, Princeton, NJ 08540

\rm
  \end{center}

\vspace{2cm}
{\bf Abstract}
\end{center}
\begin{quotation}
\noindent
Caron-Huot has recently given an interesting formula that determines OPE data in a conformal field theory in terms of a weighted integral of the four-point function over a Lorentzian region of cross-ratio space. We give a new derivation of this formula based on Wick rotation in spacetime rather than cross-ratio space. The derivation is simple in two dimensions but more involved in higher dimensions. We also derive a Lorentzian inversion formula in one dimension that sheds light on previous observations about the chaos regime in the SYK model.
\end{quotation}

\setcounter{page}{0}
\setcounter{tocdepth}{2}
\setcounter{footnote}{0}
\newpage

\tableofcontents

\pagebreak

\section{Introduction}
The operator product expansion in a conformal field theory implies that one can write a four-point correlation function as a discrete sum of conformal blocks corresponding to the physical operators of the theory:
\be\label{OPE}
\langle O_1(x_1)\cdots O_4(x_4)\rangle = \sum_{\Delta,J}p_{\Delta,J} G^{\Delta_i}_{\Delta,J}(x_i).
\ee
The conformal block $G^{\Delta_i}_{\Delta,J}(x_i)$ gives the total contribution to their four-point function coming from operators in a multiplet with a primary of dimension $\Delta$ and spin $J$. The superscript $\Delta_i$ represents dependence on the dimensions of the four external operators $O_i$. The coefficient $p_{\Delta,J}$ is a product of OPE coefficients, and the sum runs over the particular set of operators that we have in a given theory.

It is sometimes useful to think about this expansion as arising from a more primitive formula where we simply expand the four-point function in terms of a complete basis of single-valued functions $\Psi^{\Delta_i}_{\Delta,J}(x_i)$.\footnote{Such expansions can be thought of in terms of harmonic analysis on the conformal group $\SO(d+1,1)$. Harmonic analysis was first applied to conformal field theory in the 70's~\cite{MackBook1974,Mack:1974sa,Dobrev:1975ru,Dobrev:1977qv}. Recently there has been renewed interest in these methods~\cite{Hogervorst:2017sfd,Hogervorst:2017kbj,Gadde:2017sjg}, partly due to their role in the large-$N$ solution of the SYK model~\cite{Maldacena:2016hyu}.} These functions are sometimes called conformal partial waves, and they are given by conformal blocks plus ``shadow'' blocks with $\Delta \rightarrow \widetilde{\Delta}\equiv d-\Delta$,
\be\label{defpsi}
\Psi_{\De,J}^{\De_i}(x_i) = K^{\De_3,\De_4}_{\tl\De,J} G_{\De,J}^{\De_i}(x_i) + K^{\De_1,\De_2}_{\De,J} G_{\tl \De,J}^{\De_i}(x_i).
\ee
The $K$ coefficients will be given in (\ref{eq:kcoefficient}) below. A mathematically complete set of such functions (in $d>1$) consists of partial waves with integer spin and unphysical complex dimensions, $\Delta = \frac{d}{2}+ir$, where $r$ is a nonnegative real number. These are often referred to as the principal series representations.

In addition to being complete, the principal series wave functions are also orthogonal in an appropriate sense. There is a conformally-invariant pairing between $\Psi^{\Delta_i}_{\Delta,J}$ and $\Psi^{\widetilde{\Delta}_i}_{\tl\Delta',J'}$ where we simply multiply the functions and integrate over all four external points. We also must divide by the volume of the conformal group $\SO(d+1,1)$, since the resulting integral is invariant under simultaneous conformal transformations of the four points. (In practice, this means we must gauge fix and insert the appropriate Faddeev-Popov determinant.) With respect to this pairing, we have the orthogonality relation,
\be\label{innerproddef}
\Big(\Psi^{\Delta_i}_{\Delta,J},\Psi^{\widetilde{\Delta}_i}_{\widetilde{\Delta}',J'}\Big) \equiv\int \frac{d^dx_1\cdots d^d x_4}{\vol(\SO(d{+}1,1))} \Psi^{\Delta_i}_{\Delta,J}(x_i)\Psi^{\widetilde{\Delta}_i}_{\widetilde{\Delta}',J'}(x_i) = n_{\Delta,J}\,2\pi\delta(r-r')\delta_{J,J'},
\ee
where the normalization coefficient $n_{\Delta,J}$ will be given in (\ref{normcoeff}) below. Here $\Delta = \frac{d}{2}+ir$ and $\widetilde{\Delta}' = \frac{d}{2}-ir'$ and we assume $r,r'\geq 0$.

Using this set of principal series wave functions, the four point function can be written
\begin{align}\label{primitive}
\langle O_1(x_1)\cdots O_4(x_4)\rangle  &= \sum_{J=0}^\infty\int_{\frac{d}{2}}^{\frac{d}{2}+i\infty} \frac{d\Delta}{2\pi i}\, \frac{I_{\Delta,J}}{n_{\Delta,J}}\, \Psi^{\Delta_i}_{\Delta,J}(x_i) + \text{(non-norm.)} \\ &= \sum_{J=0}^\infty\int_{\frac{d}{2}-i\infty}^{\frac{d}{2}+i\infty} \frac{d\Delta}{2\pi i}\,\frac{I_{\Delta,J}}{n_{\Delta,J}}\, K^{\De_3,\De_4}_{\tl\De,J} G^{\Delta_i}_{\Delta,J}(x_i) + \text{(non-norm.)}.
\end{align}
In the first line we introduced the coefficient function $I_{\Delta,J}$, dividing by $n_{\Delta,J}$ for convenience. This function $I_{\Delta,J}$ contains all of the theory-specific information in the four point function, and it will be the focus of this paper. In the second line we inserted (\ref{defpsi}) and then absorbed the second term by extending the region of integration of the first term. The non-normalizable contributions will be discussed in appendix~\ref{app:non-norm}. 

We can now understand how to recover the OPE presentation in (\ref{OPE}): we deform the contour of integration over $\Delta$ to the right, picking up poles along the real $\Delta$ axis at the locations of physical operators. The residues are proportional to $p_{\Delta,J}$.

Often, we imagine using (\ref{OPE}) and (\ref{primitive}) to determine the four-point function in a case where we know the OPE data or expansion coefficient $I_{\Delta,J}$. However, for some applications, it is useful to imagine applying the logic in reverse. Then we assume that the four-point function (or some contribution to it) is given, and we want to evaluate the corresponding OPE or coefficient function $I_{\Delta,J}$. To do this we take the pairing of $\Psi$ with the four-point function. Using (\ref{innerproddef}) and (\ref{primitive}), we find an inversion formula\footnote{We use the notation that $O_i$ is always at position $x_i$ unless otherwise specified.}
\be\label{eucinv}
I_{\Delta,J} = \left(\langle O_1\cdots O_4\rangle,\Psi^{\widetilde{\Delta}_i}_{\widetilde{\Delta},J}\right) = \int \frac{d^dx_1\cdots d^d x_4}{\vol(\SO(d{+}1,1))} \langle O_1\cdots O_4\rangle\Psi^{\widetilde{\Delta}_i}_{\widetilde{\Delta},J}(x_i).
\ee
In this formula, all four points are integrated over $d$-dimensional Euclidean space. By partially gauge-fixing the $\SO(d{+}1,1)$ symmetry, this can be reduced to an integral over cross ratios.

We would like to emphasize that (\ref{eucinv}) is quite trivial, simply expressing the orthogonality of the partial waves. Recently, a much more interesting formula for $I_{\Delta,J}$ has been presented by Caron-Huot~\cite{Caron-Huot:2017vep}. This involves an integral over two Lorentzian regions, with an integrand given by a special type of conformal block multiplied by a double commutator, either $\langle [O_1,O_3][O_2,O_4]\rangle$ or $\langle [O_1,O_4][O_2,O_3]\rangle$, depending on the region. This formula has several advantages, such as the fact that it can be analytically continued in the spin $J$, and that for real dimension and spin the integrand satisfies positivity conditions.

The purpose of this paper is to give an alternate derivation of Caron-Huot's more interesting formula. Our strategy is as follows. We start from the formula (\ref{eucinv}), and we represent the $\Psi$ function using the shadow representation, as an integral over a fifth point. The formula for $I_{\Delta,J}$ is now a conformally-invariant integral over five points in Euclidean space. The idea is to Wick-rotate and deform the contour of integration over these points. We end up integrating over a subregion of Lorentzian spacetime such that e.g.\ $x_3$ is in the future of $x_1$ and $x_2$ is the future of $x_4$, but all other relationships between points are spacelike. After integrating out some of the coordinates using conformal symmetry, this becomes Caron-Huot's formula.

In slightly more detail, the specific Wick rotation is simplest to describe after making a partial gauge fixing of $\SO(d{+}1,1)$, where we set $x_1 = (1,0,0,\cdots)$, $x_2 = 0$, and $x_5 = \infty$. We then Wick-rotate the integral over the remaining points $x_3,x_4$. The integrand has branch point singularities at locations where $x_3$ or $x_4$ become null separated from $x_1$ or $x_2$. We deform the contour for each of $x_3,x_4$ to pick up the discontinuity across the corresponding branch cuts. For each of $x_3,x_4$, the discontinuity leads to a commutator between one of $O_3,O_4$ and $O_1,O_2$. Deforming the contour in both variables (which is valid for $J>1$) gives double commutators of the type described above, integrated over a subset of the Lorentzian space for $x_3,x_4$:
\begin{align}\label{cont1}
I_{\Delta,J} &=  -\hat C_J(1)\Bigg[\int_{3>1,2>4} \frac{d^d x_3 d^d x_4}{\vol(\SO(d{-}1))} \frac{\<[O_4,O_2][O_1,O_3]\>}{|x_{34}|^{J+2d-\De_3-\De_4-\De}} (m\cdot x_{34})^J\theta(m\cdot x_{34})\\&\hspace{20pt}+(-1)^J\int_{4>1,2>3} \frac{d^d x_3 d^d x_4}{\vol(\SO(d{-}1))} \frac{\<[O_3,O_2][O_1,O_4]\>}{|x_{34}|^{J+2d-\De_3-\De_4-\De}} (-m\cdot x_{34})^J\theta(-m\cdot x_{34})\Bigg].\notag
\end{align}
Here $m$ is the null vector $ m^\mu = (1,1,0,\dots,0)$, the second component is the time direction. The notation $i>j$ means that $x_i$ is in the future lightcone of $x_j$. In the regions where the $\theta$ step functions are nonzero, all pairs of points not indicated in the subscript to the integral are spacelike separated. The $\hat{C}_J(1)$ constant is specified in footnote~\ref{footgeg}. The fact that we have a natural analytic continuation in spin $J$ (apart from the $(-1)^J$ factor) is obvious already from (\ref{cont1}).

 As a final step, this integral can be simplified to Caron-Huot's formula (an integral over cross ratios only) by un-gauge-fixing this integral and re-gauge-fixing in a new gauge that separates the integration variables into cross ratios and everything else. The integral over everything else gives a multiple of a funny conformal block with ``dimension'' given by $J+d-1$ and ``spin'' given by $\Delta-d+1$. Concretely, 
\begin{align}
I_{\Delta,J} = \alpha_{\Delta,J} \Bigg[(-1)^J&\int_0^1\int_0^1 \frac{d\chi d\bar \chi}{(\chi \bar \chi)^d} |\chi-\b\chi|^{d-2} G^{\tl \De_i}_{J+d-1,\Delta-d+1}(\chi,\bar\chi)\frac{\<[O_3,O_2][O_1,O_4]\>}{T^{\De_i}}  \\
&+\int_{-\infty}^0\int_{-\infty}^0\frac{d\chi d\bar \chi}{(\chi \bar \chi)^d} |\chi-\b\chi|^{d-2} \hat G^{\tl \De_i}_{J+d-1,\Delta-d+1}(\chi,\bar\chi)\frac{\<[O_4,O_2][O_1,O_3]\>}{T^{\De_i}} \Bigg].\notag
\end{align}
In this expression, $T^{\Delta_i}$ is a factor of external positions that we strip off to make the four-point function depend only on the cross ratios, see (\ref{eq:defoft}). The $\alpha$ coefficient is given in (\ref{eq:eqnforalpha}). This formula is precisely Caron-Huot's inversion formula once we convert to his $c(J,\Delta)$ using
\be\label{translate}
c(J,\Delta) = \frac{I_{\Delta,J}}{n_{\Delta,J}}K^{\De_3,\De_4}_{\tl \De,J}.
\ee
Note that this translation contains a factor of $(-1)^J$.

In the rest of the paper we will spell out the details in this argument. Although each step is simple, there are several steps involved. In two dimensions some of these can be combined, and the presentation is significantly simpler. We will go through this case first.  We also present a separate derivation for the interesting case of dimension one, where lightcone coordinates are not available but Caron-Huot's formula does have a nontrivial analog,
which played a role in~\cite{Maldacena:2016hyu}.

\section{Two dimensions}\label{twod}
In this section we will derive the Lorentzian OPE inversion formula for the special case of a conformal field theory in two spacetime dimensions. We treat this case separately because some aspects are different (and simpler!) than the $d>2$ case, which we will discuss in the next section. To further simplify the analysis, we will specialize to the case where the external dimensions are equal $\De_1=\De_2=\De_3=\De_4=\De_O$. We will study general $\De_i$ when we move to higher dimensions.

In two dimensions, the conformal group $\SL(2,\R) \x \SL(2,\R)$ has two independent quadratic Casimirs, associated with the two $\SL(2,\R)$ factors. Eigenfunctions of the Casimirs are labeled by a pair of left and right weights $(h,\bar h)$, where the dimension is $\Delta = h + \bar h$ and the spin is $J = |h-\bar h|$.\footnote{Note that $\bar h$ is not in general the complex conjugate of $h$.} The eigenfunctions are given by the shadow representation
\begin{align}
\label{eq:2dshadow}
\Psi^{\De_O}_{h,\bar h}(z_i,\bar z_i) &= \frac{1}{|z_{12}|^{2\De_O}|z_{34}|^{2 \De_O}} \Psi_{h,\bar h}(z_i,\bar z_i) \nn\\
\Psi_{h,\bar h}(z_i,\bar z_i) &= \int d^2 z_5 \p{\frac{z_{12}}{z_{15}z_{25}}}^h \p{\frac{\bar z_{12}}{\bar z_{15} \bar z_{25}}}^{\bar h}\p{\frac{z_{34}}{z_{35}z_{45}}}^{1-h} \p{\frac{\bar z_{34}}{\bar z_{35} \bar z_{45}}}^{1-\bar h}.
\end{align}
As usual in the shadow representation, $\Psi_{h,\bar h}$ is an eigenvector of the Casimirs because it is a linear combination of products of three-point functions, each of which is an eigenvector of the Casimirs.
Note that the partial wave for the exchange of a symmetric traceless tensor (STT) would be $\Psi_{h,\bar h}+\Psi_{\bar h,h}$, because STT representations are reducible in 2-dimensions (when the spin $J=|h-\bar h|$ is nonzero). Thus, $\Psi_{h,\bar h}$ is not quite analogous to $\Psi^{\De_i}_{\De,J}$ in higher dimensions, which is associated to STTs. This point will be important later. Our normalization of the two dimensional shadow integral also differs from what we will define in higher dimensions by a factor of $2^J$. 

The expansion of the four-point function in terms of $\Psi_{h,\bar h}$ can be written as
\be
\langle O_1(z_1)\cdots O_4(z_4)\rangle  = \sum_{\ell=-\infty}^\infty\int_{0}^{\infty} \frac{dr}{2\pi} \,\frac{I_{h,\b h}}{n_{h,\bar h}}\, \Psi^{\Delta_O}_{h,\bar h}(z_i) + \text{(non-norm.)},
\ee
where $h = \frac{1+\ell+ir}{2}$ and $\bar h = \frac{1-\ell+ir}{2}$. The orthogonality relation for these eigenfunctions, in the sense of (\ref{innerproddef}) is~\cite{Murugan:2017eto}
\be
\left(\Psi^{\Delta_O}_{h,\bar h},\Psi^{\widetilde{\Delta}_O}_{1-h',1-\bar h'}\right) = n_{h,\bar{h}}\,2\pi\delta(r-r')\delta_{\ell,\ell'}, \hspace{20pt} n_{h,\bar{h}} = -\frac{2\pi^3}{(2h-1)(2\b h-1)}.
\ee
To extract $I_{h,\b h}$, we must take an inner product between the four-point function $\<O_1O_2O_3O_4\>$ and the partial wave $\Psi^{\tl \De_O}_{1-h,1-\bar h}$, where $\tl \De_O=2-\De_O$. On the one hand, this is given by
\begin{align}
I_{h,\b h} &= \int \frac{d^2 z_1 \cdots d^2 z_4}{\vol(\SO(3,1))} \<O_1 O_2 O_3 O_4\> \Psi^{\tl \De_O}_{1-h,1-\bar h}(z_i,\bar z_i) \label{newonehello!}\\
&= \int \frac{d^2 \chi}{|\chi|^{4-2\De_O}} \<O_1(0)O_2(\chi)O_3(1)O_4(\oo)\> \Psi_{1-h,1-\bar h}(0,\chi,1,\oo),
\label{eq:2dgaugefixing1}
\end{align}
where in the second line we have chosen the gauge $z_1=0,z_2=\chi,z_3=1,z_4=\oo$ (and we are only writing holomorphic coordinates for brevity). This integral in terms of cross-ratios $\chi,\bar \chi$ is the usual Euclidean inversion formula.

On the other hand, plugging in the shadow representation (\ref{eq:2dshadow}), we can write the integral on the RHS of (\ref{newonehello!}) as
\begin{align}
\int \frac{d^2 z_1 \cdots d^2 z_5 }{\vol(\SO(3,1))}\frac{\<O_1 O_2 O_3 O_4\>}{|z_{12}|^{4-2\De_O}|z_{34}|^{4-2\De_O}} \p{\frac{z_{12}}{z_{15}z_{25}}}^{1-h} \p{\frac{\bar z_{12}}{\bar z_{15} \bar z_{25}}}^{1-\bar h}\p{\frac{z_{34}}{z_{35}z_{45}}}^{h} \p{\frac{\bar z_{34}}{\bar z_{35} \bar z_{45}}}^{\bar h}.
\label{eq:fivepttwod}
\end{align}
As mentioned in the introduction, it is useful to partially gauge fix (\ref{eq:fivepttwod}) in a different way, where we choose $z_1=1,z_2=0,z_5=\oo$. This gives
\begin{align}
\label{eq:kernelinterpretation}
I_{h,\b h} &= \int \frac{d^2 z_3 d^2 z_4 }{|z_{34}|^{4-2\De_O}}\<O_1 O_2 O_3 O_4\>z_{34}^h\bar z_{34}^{\bar h}.
\end{align}

Although (\ref{eq:kernelinterpretation}) treats the operators less symmetrically than (\ref{eq:2dgaugefixing1}), it is natural from a different point of view. We can think about the four-point function as a kernel that maps functions of $z_3,z_4$ to functions of $z_1,z_2$, by integrating over $z_3,z_4$. By global conformal invariance, this kernel commutes with the conformal Casimirs, so eigenfunctions of the Casimirs (like $z_{34}^h \bar z_{34}^{\bar h}$) should also be eigenfunctions of the four-point function. We could have taken (\ref{eq:kernelinterpretation}) as our starting point for the definition of $I_{h,\bar h}$. In this case, we could return to the integral over cross-ratios (\ref{eq:2dshadow}) by making the change of variables
\be
\chi = \frac{z_{34}}{(z_3-1)z_4},
\ee
and integrating over $z_4$. 

An important point is that (\ref{eq:2dgaugefixing1}) and (\ref{eq:kernelinterpretation}) only make sense if the spin $J$ is an integer, because otherwise the functions $\Psi_{1-h,1-\bar h}$ and $z_{34}^h \bar z_{34}^{\bar h}$ would not be single-valued in Euclidean signature.

\subsection{Wick rotation and the double commutator}
\label{sec:contourmanipulation}

We will now derive a different formula for $I_{h,\bar h}$ by doing the integral over $z_3,z_4$ in (\ref{eq:kernelinterpretation}) in Lorentzian signature. To Wick rotate, we use the normal Feynman continuation so that we take $\tau = (i+\epsilon)t$. Then
\be
|z|^2 = x^2+\tau^2 = x^2-t^2+i\epsilon = uv + i\epsilon.
\ee
Here we have defined $u = x-t$ and $v = x+t$. With this $i\epsilon$ prescription, the integral over Lorentzian kinematics gives the same answer as the original Euclidean integral. Our integral becomes
\begin{align}
\label{eq:afterWick2d}
I_{h,\b h} &= -\frac{1}{4} \int \frac{du_3 dv_3 du_4 dv_4}{(u_{34} v_{34})^{2-\De_O}}\<O_1 O_2 O_3 O_4\>u_{34}^h v_{34}^{\bar h},
\end{align}
where the factor of $-\frac{1}{4}$ arises because $d^2z \equiv d\tau dx = \frac{i}{2}dudv$.

It will be important to understand the locations of singularities in the complex $u,v$ planes. {In two dimensions, singularities in the four-point function only occur when some pair of external operators become null separated~\cite{Maldacena:2015iua} (in higher dimensions other singularities are possible, but they will not interfere with the analogous argument).} Since we are fixing the locations of $u_1,v_1 = 1,1$ and $u_2,v_2 = 0,0$, singularities occur when one of the following hold:
\begin{align}
u_3v_3+i\epsilon = 0, \hspace{20pt} u_4v_4 + i\epsilon = 0, \hspace{20pt} (1-u_3)(1-v_3) +i\epsilon = 0,\\
(1-u_4)(1-v_4) + i\epsilon= 0, \hspace{20pt} (u_3-u_4)(v_3-v_4) + i\epsilon = 0. 
\end{align}
Let us think about fixing $u_3,u_4$ and doing the integral over the $v_3,v_4$ variables. 
Suppose that $h=\frac{\De+J}{2}$, $\bar h = \frac{\De-J}{2}$ with $J$ positive. (If $J$ is negative, we reverse the roles of $u,v$ in the following.) For sufficiently large $J$ {(see appendix~\ref{app:reqJgtr1})}, the factor $v_{34}^{\b h}$ causes the integrand to die at large $v_3,v_4$. Thus, we can deform the $v_3,v_4$ integrals away from the real axis without worrying about contributions near infinity.

For each of the $v$ variables there are three singularities.
If all of the singularities in $v_3$ or $v_4$ are in the upper or lower half planes, then the integral will vanish. This happens if $u_{31},u_{32},u_{34}$ all have the same sign, or if $u_{41},u_{42},u_{43}$ all have the same sign.

To get a nonvanishing result, we must have one singularity on one side of the axis and two on the other side, for each of $v_3$ and $v_4$. This requires $0<u_3,u_4<1$.  We can then deform each of the $v$ contours towards the half-plane with only one singularity. This singularity is a branch point, and we can take the branch cut to lie just above or just below the real axis. The $v$ integrals then take the discontinuities across these branch cuts, which are the same as the commutators of certain pairs of operators.

\begin{figure}
\begin{center}
\includegraphics[width=\textwidth]{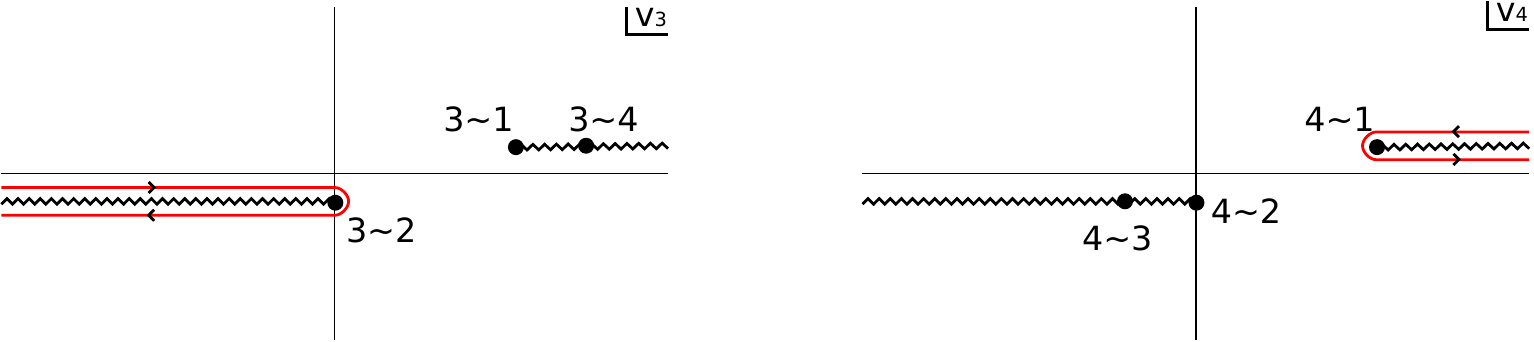}
\caption{\small{The continuation of $v_3,v_4$ in the case where $0<u_3<u_4<1$. We begin by integrating both variables over the real axis. We deform the $v_3$ contour in the lower half-plane to pick up the discontinuity across the branch cut associated with the $3\sim 2$ singularity, giving the $[O_3,O_2]$ commutator. We deform the $v_4$ contour in the upper half-plane and pick up the $[O_4,O_1]$ singularity.}}\label{fig:contourDef}
\end{center}
\end{figure}
\begin{figure}
\begin{center}
\includegraphics[width=.8\textwidth]{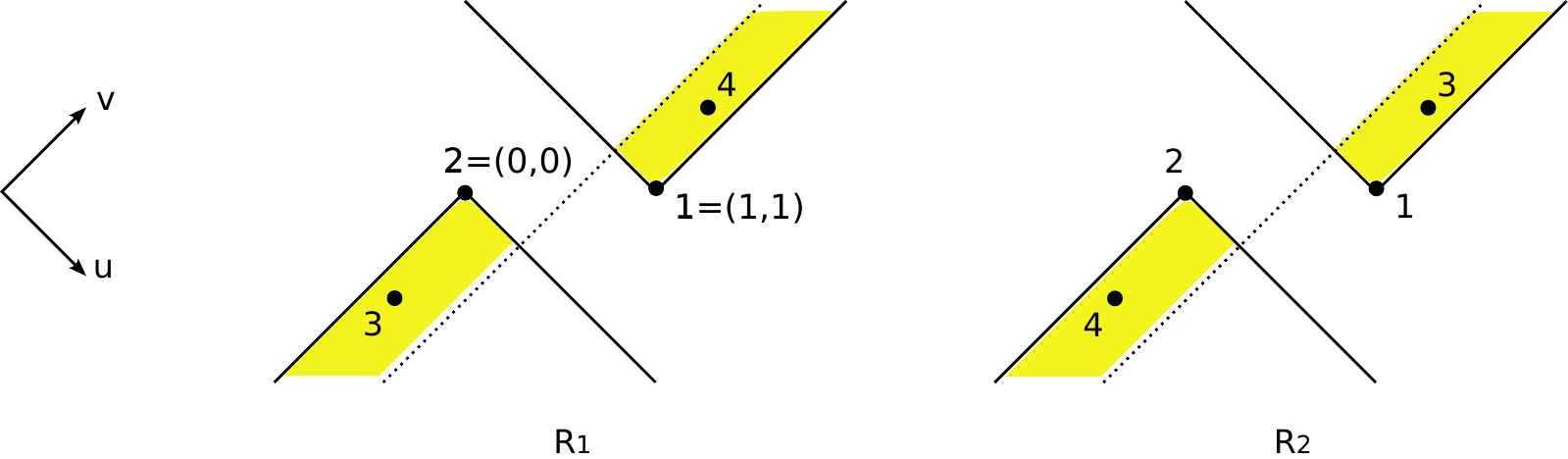}
\caption{We show typical configurations for points 3 and 4 within regions $R_1$ and $R_2$. The dotted line is not fixed in place, it is only to emphasize that points 3 and 4 must be spacelike separated. Time goes up.}\label{fig:regionsR1R2}
\end{center}
\end{figure}

For example, when $0<u_3<u_4<1$ (see figure~\ref{fig:contourDef}), we deform the $v_3$ contour towards the lower half-plane around the singularity  $v_{32}=-i\epsilon/u_{32}$ to produce a commutator $[O_3,O_2]$. Similarly, we deform the $v_4$ contour towards the upper half-plane around the singularity $v_{41}=-i\epsilon/u_{41}$ to produce $[O_1,O_4]$. In the other case $0<u_4<u_3<1$, we obtain the commutators $[O_4,O_2][O_1,O_3]$.
The precise formula we find is
\begin{align}
I_{h,\b h}=  -\frac{(-1)^J}{4}&\int_{R_1} \frac{du_3dv_3du_4dv_4}{(u_{34}v_{34})^{2-\De_O}}\langle [O_3,O_2][O_1,O_4]\rangle u_{43}^h v_{43}^{\bar h} \nn\\
&-\frac{1}{4}\int_{R_2}\frac{du_3dv_3du_4dv_4}{(u_{34}v_{34})^{2-\De_O}} \langle [O_4,O_2][O_1,O_3]\rangle u_{34}^h v_{34}^{\bar h},
\end{align}
where the two integration regions are defined by
\begin{align}
R_1:&\quad v_3<0,\quad v_4>1,\quad 0<u_3<u_4<1, \nn\\
R_2:&\quad v_3>1,\quad v_4<0,\quad 0<u_4<u_3<1.
\end{align}
The factor of $(-1)^J$ comes from writing $z_{34}^h\bar z_{34}^{\bar h} \rightarrow (-1)^J u_{43}^h v_{43}^{\bar h}$. One way to summarize the regions $R_1,R_2$ is that the operators in the commutators are timelike separated and all other pairs are spacelike separated, see figure~\ref{fig:regionsR1R2}.

Note that after our contour deformation, the integrals above can be analytically continued in $J$. For example, the first integral is over a Lorentzian region where $u_{43}$ and $v_{43}$ are real and positive, so there is no issue with single-valuedness. The factor $(-1)^J$ means that it is natural to analytically continue $C(h,\bar h)$ separately for even and odd $J$.

\subsection{Rewriting in terms of cross-ratios}

To make contact with Caron-Huot's formula, we would like to use the fact that the four-point function (and the commutators) depend only on the cross ratios. Given that $u_1=v_1 = 1$ and $u_2 = v_2 = 0$, these reduce to
\be
\label{eq:separation2d}
\chi = \frac{u_{34}}{(u_3-1)u_4}, \hspace{20pt} \bar{\chi} = \frac{v_{34}}{(v_3-1)v_4}.
\ee
 We can solve these equations for $u_3,v_3$ and change variables in the integral, so that we have an integral over $u_4,v_4,\chi,\b\chi$. Because the four-point function depends only on $\chi,\b\chi$, we can then do the integral over $u_4,v_4$ explicitly, getting exprssions involving the $SL(2,\R)$ conformal block
\begin{align}
k_{2h}(\chi) \equiv \chi^h {}_2F_1(h,h,2h,\chi),\quad \hat{k}_{2h}(\chi) \equiv (-\chi)^h {}_2F_1(h,h,2h,\chi).
\end{align}
The final answer one finds is 
\begin{align}
I_{h,\b h} &= -\frac{1}{4}\frac{\Gamma(h)^2}{\Gamma(2h)}\frac{\Gamma(1{-}{\bar h})^2}{\Gamma(2{-}2{\bar h})}\Bigg[(-1)^J\int_0^1\int_0^1\frac{d\chi d\b \chi}{(\chi\b \chi)^{2-\De_O}}\langle [O_3,O_2][O_1,O_4]\rangle k_{2h}(\chi) k_{2(1-\bar h)}(\b\chi)\notag\\ &\hspace{40pt}+\int_{-\infty}^0\int_{-\infty}^0\frac{d\chi d\b \chi}{(\chi\b \chi)^{2-\De_O}}\langle [O_4,O_2][O_1,O_3]\rangle \hat{k}_{2h}(\chi)\hat{k}_{2(1-\b h)}(\b \chi)\Bigg].
\end{align}
One can check that this formula agrees with~\cite{Caron-Huot:2017vep} once we translate using (\ref{translate}) which for this special case reads
\be
c(J,\Delta) = \frac{(-1)^J}{2\pi^2}\frac{\Gamma(h)^2}{\Gamma(2h{-}1)}\frac{\Gamma(2{-}2\b h)}{\Gamma(1{-}\b h)^2}\, I_{h,\b h}.
\ee
where $J = h-\b h$ and $\Delta = h+\b h$.

\section{Higher dimensions}

Our discussion in higher dimensions will mirror the one in two dimensions, but with some new complications. Firstly, note that our two-dimensional derivation required a choice that depended on the sign of
$h - \bar h $. However, the partial wave for a symmetric traceless tensor (STT) contains two terms with the role of $h$ and $\bar h$ swapped: $\Psi_{h,\bar h} + \Psi_{\bar h, h}$. If we take an inner product of $\<O_1O_2O_3O_4\>$ with a STT partial wave, we obtain (\ref{eq:afterWick2d}) with $u_{34}^h v_{34}^{\bar h}$ replaced by $u_{34}^h v_{34}^{\bar h}+u_{34}^{\bar h} v_{34}^{h}$. These two terms must be treated separately: for the first term, we must deform the $v$ contour for fixed $u$, and for the second term we must deform the $u$ contour for fixed $v$. In the previous section, we avoided this complication by only discussing the ``chiral half" of a partial wave. However, in higher dimensions, the complication is unavoidable because operators transform as STTs. Our approach will be to isolate an individual null direction (similarly to isolating one of the two terms above), and perform the two-dimensional contour manipulation for that null direction.

The second complication is that in higher-dimensions, after Wick rotation to Lorentzian signature and performing a contour manipulation to obtain the double-commutator, the separation of variables into cross-ratios and non-cross-ratios (as in (\ref{eq:separation2d})) is more difficult. To do this, we will un-isolate the null directions by integrating over them. The result can then be re-interpreted as a gauge-fixed five-point integral, this time in Lorentzian signature. Choosing a different gauge, we obtain an integral over cross ratios that reproduces Caron-Huot's formula.

To summarize, the logical outline of our derivation is as follows:
\begin{enumerate}
\item Set up the inner product between the four-point function and a partial wave $\Psi^{\De_i}_{\De,J}$ as an integral over five Euclidean points, with $x_5$ being the point we integrate over in the shadow representation of $\Psi$.
\item Choose the gauge $x_1=(1,0,\dots,0)$, $x_2=(0,\dots,0)$, $x_5=\oo$.
\item Isolate a single null direction (using a particular representation of Gegenbauer polynomials discussed below), and Wick rotate to Lorentzian signature.
\item Perform the two-dimensional contour deformation from section~\ref{sec:contourmanipulation} to obtain a double commutator and an integral over a restricted Lorentzian region.
\item Integrate over null directions.
\item Un-gauge-fix the integral and then re-gauge-fix in a different gauge that separates the integration variables into cross-ratios plus non-cross-ratio degrees of freedom.
\item Evaluate the integral over non-cross-ratio degrees of freedom in the limit of small cross-ratios. This fixes the integral for all values of the cross ratios because we know it has to give an eigenfunction of the conformal Casimir.
\end{enumerate}

\subsection{Initial setup and gauge fixing}

With these preliminaries out of the way, our first task is to write the inner product between the partial wave and our four-point function as a conformally-invariant integral over five points. The fifth point arises from the shadow representation of the partial wave, which in general dimensions has the form:
\be\label{shadow}
\Psi^{\Delta_i}_{\De,J}(x_i) = \int d^dx_5 \< O_1  O_2 O_5^{\mu_1\cdots \mu_J}\>\< \tl O_{5,\mu_1\cdots\mu_J} O_3 O_4\>.
\ee
Here the three-point functions are given by e.g.\footnote{When we write a two- or three-point function, we mean a conformally-invariant structure with the given quantum numbers (with a simple normalization that we specify). In particular, three-point functions don't include OPE coefficients. By contrast, the four-point function $\<O_1 O_2 O_3 O_4\>$ can be thought of as a physical correlation function in some  theory.}
\be
\<O_1 O_2 O_5^{\mu_1 \cdots \mu_J}\> =\frac{Z^{\mu_1} \cdots Z^{\mu_J} - \textrm{traces}}{|x_{12}|^{\De_1+\De_2-\De}|x_{15}|^{\De_1+\De-\De_2}|x_{25}|^{\De_2+\De-\De_1}}, \hspace{20pt} 
Z^\mu \equiv \frac{|x_{15}||x_{25}|}{|x_{12}|}\p{\frac{x^\mu_{15}}{x_{15}^2}- \frac{x^\mu_{25}}{x_{25}^2}}.
\ee
This leads to the explicit formula for the partial wave
\begin{align}\label{shadowgend}
\Psi^{\Delta_i}_{\De,J}(x_i) =&\int d^dx_5 \frac{1}{|x_{12}|^{\De_1+\De_2-\De}|x_{15}|^{\De_1+\De-\De_2}|x_{25}|^{\De_2+\De-\De_1}}\nn\\ &\hspace{30pt}\x\frac{1}{|x_{34}|^{\De_3+\De_4 - \tl\De}|x_{35}|^{ \De_3+ \tl\De- \De_4}|x_{45}|^{\De_4+ \tl\De -  \De_3}}\hat{C}_J(\eta),
\end{align}
where we have defined the conformal invariant
\be\label{defeta}
\eta = \frac{|x_{15}||x_{25}|}{|x_{12}|}\frac{|x_{35}||x_{45}|}{|x_{34}|}\left(\frac{\vec{x}_{15}}{x_{15}^2}-\frac{\vec{x}_{25}}{x_{25}^2}\right)\cdot \left(\frac{\vec{x}_{35}}{x_{35}^2} - \frac{\vec{x}_{45}}{x_{45}^2}\right),
\ee
and we wrote the sum over polarizations in terms of a Gegenbauer polynomial\footnote{\label{footgeg}We define  $\hat{C}_J(x) \equiv \frac{\G(J+1)\G(\frac{d-2}{2})}{2^J \G(J+\frac{d-2}{2})}C_J^{d/2-1}(x)$ where $C_J^{d/2-1}(x) \equiv \frac{\Gamma(J+d-2)}{\Gamma(J+1)\Gamma(d-2)}{}_2F_1(-J,J+d-2,\frac{d-1}{2},\frac{1-x}{2})$.} using
\be
|n|^J|m|^J\hat{C}_J\p{\frac{n\.m}{|n||m|}} = (n^{\mu_1}\cdots n^{\mu_J}-\textrm{traces})(m_{\mu_1}\cdots m_{\mu_J}-\textrm{traces}).
\ee
Note that $\hat{C}_J(x)$ is normalized so that the coefficient of $x^J$ is one. 

The Euclidean inversion formula (\ref{eucinv}) is an inner product between our four-point function and the partial wave $\Psi^{\tl\De_i}_{\tl\De,J}$ where we replace all operators by their shadows $\tl\Delta = d-\Delta$. Using the shadow representation of this partial wave, (\ref{eucinv}) becomes
\be
\label{eq:fivepointintegral}
I_{\Delta,J} = \int \frac{d^d x_1 \cdots d^d x_5}{\vol(\SO(d{+}1,1))} \<O_1O_2O_3O_4\> \<\tl O_1 \tl O_2 \tl O_5^{\mu_1\cdots \mu_J}\>\< O_{5,\mu_1\cdots\mu_J}\tl O_3\tl O_4\>.
\ee
This is a conformally-invariant integral. As in the two-dimensional case, it will be helpful to partially fix the gauge for the conformal group by setting $x_5=\oo$, $x_1=(1,0,\dots,0)$ and $x_2=(0,\dots,0)$.  
We can define $\vol(\SO(d+1,1))$ so that gauge-fixing three points to $0,1,\oo$ gives a Faddeev-Popov determinant of 1. The above formula then becomes
\be
I_{\Delta,J}= \int \frac{d^d x_3 d^d x_4}{\vol(\SO(d{-}1))}  \frac{\<O_1 O_2 O_3 O_4\>}{|x_{34}|^{2d-\De_3-\De_4-\De}} \hat C_J\p{\frac{x_{34}\.x_{12}}{|x_{34}||x_{12}|}},
\label{eq:kernelformulaforc}
\ee
where $\SO(d{-}1)$ is the stabilizer group of three fixed points. Our convention for the measure on $\SO(n)$ is that a $2\pi$-rotation should have length $2\pi$. This gives
\begin{align}
\vol(\SO(n)) &= \vol(S^{n-1})\vol(\SO(n{-}1)).
\label{eq:volumeratios}
\end{align}

\subsection{Isolating a null direction}
\label{sec:isolatingnull}

We cannot perform our contour manipulation with (\ref{eq:kernelformulaforc}) because for large $J$, the Gegenbauer polynomial $\hat C_J\p{\frac{x_{34}\.x_{12}}{|x_{34}||x_{12}|}}$ grows in every null direction. Instead, we would like to find an integrand that does not grow along some null direction.

Consider the following representation of the Gegenbauer polynomial:
\begin{align}
\label{eq:gegenbauerrep}
|x|^J \hat C_J\p{\frac{x^0}{|x|}} &= \frac{\hat C_J(1)}{\vol(S^{d-2})} \int_{S^{d-2}} d^{d-2} \hat e\, (n\.x)^J,
\end{align}
where $\hat e$ is a unit vector in $d-1$ dimensions, the integral is over the $d-2$ sphere, and $n=(1,i\hat e)$ is a null vector. Because $n$ is null, the right-hand side is a harmonic polynomial of degree $J$ in $x$ (and thus it transforms as a traceless symmetric tensor of spin $J$). It is a function of $x^0$ and $|x|$ alone because it involves an average over transverse rotations. These conditions uniquely specify the Gegenbauer polynomial up to some constant, which we have fixed out front.\footnote{One way to understand why the $(d{-}2)$-dimensional integral (\ref{eq:gegenbauerrep}) gives a natural object in $d$-dimensions is as follows. After Wick rotating $x^0 \to i x^0$ and redefining $n\to -i n$ (note that this is {\it not\/} the Wick rotation we do in section~\ref{sec:wickhigherd}), the integral (\ref{eq:gegenbauerrep}) becomes a manifestly $\SO(d-1,1)$-invariant integral over the projective null-cone in $d$-dimensions:
\begin{align}
\label{eq:conformalintegral}
|x|^J |y|^{2-d-J} \hat C_J\p{\frac{x\.y}{|x||y|}} &\propto \frac{1}{\vol(\R^+)}\int d^d n\, \de(n^2) \theta(n^0) (n\.x)^J (n\.y)^{2-d-J},
\end{align}
where $y=(1,0,\dots,0)$.
 Integrals of exactly the same type in $(d{+}2)$-dimensions appeared in~\cite{SimmonsDuffin:2012uy}, where they are helpful for understanding the shadow representation of conformal blocks.}

Plugging (\ref{eq:gegenbauerrep}) into (\ref{eq:kernelformulaforc}) gives
\begin{align}
I_{\Delta,J} &= \frac{\hat C_J(1)}{\vol(S^{d-2})}\int \frac{d^d x_3 d^d x_4}{\vol(\SO(d{-}1))} \int_{S^{d-2}} d\hat e \frac{\<O_1 O_2 O_3 O_4\>}{|x_{34}|^{J+2d-\De_3-\De_4-\De}} (x_{34}^0 + i \hat e\.\vec x_{34})^J.
\end{align}
In this formula, we are averaging over rotations that fix $x_{12}$. However, the four-point function is invariant under such rotations, so the answer is given by fixing $\hat e$ to a unit vector of our choice and multiplying by $\vol(S^{d-2})$. For example, let us choose $\hat e = (0,1,0,\dots,0)$, giving
\begin{align}
\label{eq:lightconeisolated}
I_{\Delta,J} &= \hat C_J(1)\int \frac{d^d x_3 d^d x_4}{\vol(\SO(d{-}1))} \frac{\<O_1 O_2 O_3 O_4\>}{|x_{34}|^{J+2d-\De_3-\De_4-\De}} (x_{34}^0 + i x_{34}^1)^J.
\end{align}
Equation (\ref{eq:lightconeisolated}) is now completely analogous to (\ref{eq:kernelinterpretation}) in the 2d case.

\subsection{A shortcut (optional)}\label{shortcut}

A simpler way to arrive at (\ref{eq:lightconeisolated}) is to think of the four-point function as a kernel taking functions of $x_{3,4}$ to functions of $x_{1,2}$ by integration over $x_{3,4}$. As discussed in the 2d case, this kernel commutes with the conformal Casimirs, and hence they can be simultaneously diagonalized. Consider the eigenvector $\<\tl O_3 \tl O_4 O_5\>$ where $O_5$ has dimension $\De$ and spin $J$. Let the eigenvalue of the four-point function be $k_{\De,J}$,
\begin{align}
\label{eq:eigenvalueeq}
k_{\De,J}\<O_1 O_2 O^{\mu_1\cdots \mu_J}_5\> &= \int d^d x_3 d^d x_4 \<O_1O_2O_3O_4\>\<\tl O_3 \tl O_4 O_5^{\mu_1\cdots \mu_J}\>.
\end{align}
We can relate $k_{\De,J}$ to $I_{\Delta,J}$ by taking an inner product of both sides with the shadow three-point function $\<\tl O_1 \tl O_2 \tl O_5\>$,
\begin{align}
&k_{\De,J} \int \frac{dx_1 dx_2 dx_5}{\vol(\SO(d{+}1,1))} \<O_1O_2O_5^{\mu_1\cdots\mu_J}\>\<\tl O_1 \tl O_2 \tl O_{5,\mu_1\cdots\mu_J}\> \nn\\
 &= \int \frac{dx_1\cdots dx_5}{\vol(\SO(d{+}1,1))}\<O_1O_2O_3O_4\>\<\tl O_1 \tl O_2 \tl O_{5,\mu_1\cdots\mu_J}\>\<O^{\mu_1\cdots \mu_J}_5\tl O_3 \tl O_4 \> \nn\\
&= I_{\Delta,J}.
\end{align}
The constant on the left-hand side can be computed
by gauge fixing $x_1=0,x_2=e,x_5=\oo$ for some unit vector $e$,
\begin{align}
&\int \frac{dx_1 dx_2 dx_5}{\vol(\SO(d{+}1,1))} \<O_1O_2O_5^{\mu_1\cdots\mu_J}\>\<\tl O_1 \tl O_2 \tl O_{5,\mu_1\cdots\mu_J}\>\nn\\
&=
\frac{1}{\vol(\SO(d{-}1))} \<O_1(0)O_2(e)O_5^{\mu_1\cdots\mu_J}(\oo)\>\<\tl O_1(0) \tl O_2(e) \tl O_{5,\mu_1\cdots\mu_J}(\oo)\> \nn\\
&= \frac{\hat C_J(1)}{\vol(\SO(d{-}1))}.
\end{align}
Thus
\begin{align}
k_{\De,J} \frac{\hat C_J(1)}{\vol(\SO(d{-}1))} &= I_{\Delta,J}.
\end{align}
Now we can set $x_5=\oo$, $x_1=(1,0,\dots,0)$ and $x_2=(0,\dots,0)$ in (\ref{eq:eigenvalueeq}) and contract with a null vector $n=(1,i,0,\dots,0)$, to obtain (\ref{eq:lightconeisolated}).

This approach avoids the special formula (\ref{eq:gegenbauerrep}) and makes it immediately clear why only one null direction matters. On the other hand, the discussion in section~\ref{sec:isolatingnull} shows us how to go back from (\ref{eq:lightconeisolated}) to the more symmetrical five-point integral (\ref{eq:fivepointintegral}): we must average over null-directions and then un-gauge-fix the five-point integral. A similar procedure will be useful in Lorentzian signature in the next section.

\subsection{Wick rotation and the double commutator}
\label{sec:wickhigherd}

We now return to the derivation. The next step is to Wick rotate the integral (\ref{eq:lightconeisolated}) by setting $x^1 = it$. Note that we are Wick-rotating the second coordinate in the list $(x^0,x^1,x^2,\cdots)$. The integral is then
\begin{align}
I_{\Delta,J} &= -\hat C_J(1)\int \frac{d^d x_3 d^d x_4}{\vol(\SO(d{-}1))} \frac{\<O_1 O_2 O_3 O_4\>}{(x_{34}^2)^{\frac{J+2d-\De_3-\De_4-\De}{2}}} u_{34}^J.
\end{align}
where $u = x^0+ix^1 = x^0-t$. The $d^dx$ measures are now assumed to be in Lorentzian signature $d^dx = dx^0dtdx^2\cdots dx^{d-1}$, and we have an overall minus sign from two Wick rotations $dx^1 = idt$.
The Feynman $i\epsilon$ is understood in the denominator.

One can now follow the same contour deformation strategy that we discussed in the two-dimensional case. The extra spatial coordinates affect the locations of the singularities, but not the half-plane that they lie in, so the contour deformation argument is the same: the $v_3$ or $v_4$ contours can be deformed to give zero unless $0<u_3,u_4<1$. The two cases $u_3<u_4$ and $u_4<u_3$ have to be treated separately, and as before each reduces to a double-commutator, but now integrated within the past and future $d$-dimensional lightcones of points 1 and 2. Introducing a null vector $m^\mu = (1,1,0,\dots,0)$ so that $m\cdot x = u$, the answer can be written as
\begin{align}\label{afterdeform}
I_{\Delta,J} &=  -\hat C_J(1)\Bigg[(-1)^J\int_{4>1,2>3} \frac{d^d x_3 d^d x_4}{\vol(\SO(d{-}1))} \frac{\<[O_3,O_2][O_1,O_4]\>}{|x_{34}|^{J+2d-\De_3-\De_4-\De}} (-m\cdot x_{34})^J\theta(-m\cdot x_{34})\notag\\&\hspace{60pt}+\int_{3>1,2>4} \frac{d^d x_3 d^d x_4}{\vol(\SO(d{-}1))} \frac{\<[O_4,O_2][O_1,O_3]\>}{|x_{34}|^{J+2d-\De_3-\De_4-\De}} (m\cdot x_{34})^J\theta(m\cdot x_{34})\Bigg],
\end{align}
where $i>j$ in the subscript of the integral means that $x^i$ is confined to the future lightcone of $x_j$. Note that the first and second lines are related to each other by a factor of $(-1)^J$ and interchanging $3\leftrightarrow 4$. Because the interval $x_{34}$ is now constrained to be spacelike, we have safely replaced $(x_{34}^2)^{1/2} \to |x_{34}|$. {In appendix~\ref{app:reqJgtr1} we give a more careful justification of the contour deformation, concluding as in~\cite{Caron-Huot:2017vep} that it is valid if $J>1$, so (\ref{afterdeform}) should be understood as correct for $J \ge 2$.}

If we substitute an individual block (or partial wave) in the $12\to 34$ channel into the double commutator, the result vanishes. We can understand this by thinking about the shadow representation
\begin{align}
\label{eq:shadowrepofblock}
\Psi &\sim \int d^d x_5 \<O_1 O_2 O_5\>\<\tl O_5 O_3 O_4\>,
\end{align}
where we Wick rotate $x_5$ to Lorentzian signature.
A nonzero commutator $[O_3,O_2]$ requires a singularity when $O_3$ and $O_2$ are lightlike separated. Although the integrand has no such singularities, the integral can have a singularity coming from the regime where $x_5$ is lightlike separated from $x_2$ and $x_3$. However, generically $x_5$ cannot be simultaneously lightlike separated from $O_2 O_3$ and  $O_1 O_4$. This is possible for special configurations of $O_1\cdots O_4$, but the singularities associated with such  configurations can be avoided when computing discontinuities. Hence, the double commutator $[O_2,O_3][O_1,O_4]$ vanishes in (\ref{eq:shadowrepofblock}). A nonzero contribution to (\ref{afterdeform}) only comes about because of an infinite sum over blocks, which produces new singularities.

\subsection{Averaging over null directions}
From this point forward in the derivation, the goal is to reduce (\ref{afterdeform}) to an integral over cross ratios. A first step is to average over our arbitrary choice of a null vector. We can do this by applying a transformation $g\in \SO(d{-}2,1)$ to our vector $m$, where $g$ acts trivially on $m^{0}$ and as a Lorentz transformation on the remaining $(d{-}1)$ components.

Averaging $g$ over $\SO(d{-}2,1)$ in e.g.\ the second line in (\ref{afterdeform}) becomes\footnote{When we write an indefinite orthogonal group $\SO(p,q)$, we always mean the connected component of the identity in that group.}
\begin{align}\label{intoverg}
\frac{-\hat C_J(1)}{\vol(\SO(d{-}2,1))}&\int_{3>1,2>4} \frac{d^d x_3 d^d x_4}{\vol(\SO(d{-}1))} \frac{\<[O_4,O_2][O_1,O_3]\>}{|x_{34}|^{J+2d-\De_3-\De_4-\De}}\int_{\SO(d-2,1)}dg (gm\cdot x_{34})^J\theta(gm\cdot x_{34}).
\end{align}
This expression looks ill-defined, since the volume of $\SO(d-2,1)$ is infinite. However, after integrating over $g$, the integrand of the $x_3,x_4$ integral is $\SO(d-2,1)$-invariant, and therefore divergent in a way that cancels this factor. If we like, dividing by $\vol(\SO(d-2,1))$ can be implemented by gauge-fixing the integral over $x_3,x_4$.

The integral over $g$ in (\ref{intoverg}) will give some solution to the Gegenbauer differential equation, but it will no longer be a polynomial. To find out what function we get, we can use a $\SO(d{-}2,1)$ transformation to set $x_{34}=x=(x^0,x^1,0,\dots,0)$ (with $x^1<x^0$ so that $x$ is spacelike) and evaluate
\begin{align}
\int_{\SO(d{-}2,1)} dg (gm\.x)^J \theta(gm\.x) &=\vol(\SO(d{-}2))\vol(S^{d-3}) \int_{0}^{\text{arccosh}\frac{x^0}{x^1}} d\beta\, \left(\sinh\beta\right)^{d-3} \left(x^0-x^1 \cosh \beta\right)^J 
\nn\\
&=\vol(\SO(d{-}2))|x|^J B_J\p{\frac{x^0}{|x|}}.
\label{eq:bcalculation}
\end{align}
{The function $B_J(y)$ can be determined exactly\footnote{{After changing variables to $z = \cosh\beta$, the integral becomes a standard hypergeometric integral. A form that makes the large $y$ behavior and the branch cut between $-1$ and $1$ obvious can be given after making a couple of quadratic transformations of the resulting hypergeometric function:
\be\label{eq:defofBJ}
B_J(y)\equiv \frac{\pi^{\frac{d-2}{2}}\G(J{+}1)}{2^{J}\G(J{+}\frac d 2)}(1+y)^{2-d-J}{}_2F_1\p{J+\frac{d{-}1}{2},J+d-2,2J+d-1,\frac{2}{1+y}}.
\ee
Also, note that in $d = 3$ dimensions, $B_J(y)$ is a Legendre $Q$-function.}}; however the only property that we will need is that for large $y$ it behaves as 
\be\label{largeyB}
B_J(y) \sim \frac{\pi^{\frac{d-2}{2}}\Gamma(J+1)}{2^J\Gamma(J+\frac{d}{2})}y^{2-d-J} \hspace{20pt} |y| \gg 1.
\ee
This is easy to see from the integral in (\ref{eq:bcalculation}), taking $x^1$ close to $x^0$ so that $|x|$ is small, and doing the integral for small $\beta$.}

Using (\ref{eq:bcalculation}), our formula (\ref{afterdeform}) can therefore be written as
\begin{align}\label{averaged}
I_{\Delta,J} &=  -\frac{\hat C_J(1)}{\vol(S^{d-2})}\Bigg[(-1)^J\int_{4>1,2>3} \frac{d^d x_3 d^d x_4}{\vol(\SO(d{-}2,1))} \frac{\<[O_3,O_2][O_1,O_4]\>}{|x_{34}|^{2d-\De_3-\De_4-\De}} B_J(-\eta) \notag\\&\hspace{80pt}+\int_{3>1,2>4} \frac{d^d x_3 d^d x_4}{\vol(\SO(d{-}2,1))} \frac{\<[O_4,O_2][O_1,O_3]\>}{|x_{34}|^{2d-\De_3-\De_4-\De}} B_J(\eta)\Bigg].
\end{align}
Here $\eta = \frac{x_{12}\cdot x_{34}}{|x_{12}||x_{34}|}$. Note that for the configuration in the first line $\eta<0$ and for the configuration in the second line, $\eta >0$, so in both cases the argument of $B_J$ is positive.

\subsection{Changing gauge}
At this point, we would like to separate the integration variables $x_3,x_4$ into two cross ratios $\chi,\b\chi$ and everything else, and then do the integral over everything else once and for all. In practice, it is convenient to do this by recognizing (\ref{averaged}) as a gauge-fixed version of a conformally-invariant integral over five points, and then fixing the gauge in a different way where the cross ratios are manifest. 

In this section, we will work with the contribution to $I_{\Delta,J}$ on the first line in (\ref{averaged}), adding in the second line at the end. The un-gauge fixed version of this contribution is
\begin{align}
I_{\Delta,J} &\supset-\frac{\hat C_J(1)}{\vol(S^{d-2})}(-1)^J\int_{4>1,2>3} \frac{d^d x_1 \cdots d^d x_5}{\vol(\SO(d,2))}  \frac{\<[O_3,O_2][O_1,O_4]\>}{|x_{12}|^{\tl\De_1+\tl\De_2-\tl\De}|x_{15}|^{\tl\De_1+\tl\De-\tl\De_2}|x_{25}|^{\tl\De_2+\tl\De-\tl\De_1}}\notag\\ &\hspace{100pt}\x\frac{B_J(-\eta)}{|x_{34}|^{\tl\De_3+\tl\De_4 - \De}|x_{35}|^{ \tl\De_3+ \De- \tl\De_4}|x_{45}|^{\tl\De_4+ \De -  \tl\De_3}}.
\end{align}
Here the integral is over configurations such that, apart from the two timelike relations described in the subscript of the integral, all pairs of points are spacelike separated. In this expression, $\eta$ is defined as in (\ref{defeta}), with a dot product taken in Lorentzian signature. 

The variables $x_1\cdots x_5$ should be understood as coordinates on the conformal completion of Minkowski space, i.e.\ the Lorentzian cylinder $S^{d-1}\times \R$. If we partially gauge-fix by fixing the location of $x_5$, then the condition that all other points should be spacelike separated from $x_5$ forces $x_1\cdots x_4$ to be in a single Minkowski diamond of the cylinder. In the natural Minkowski space coordinates on this patch, $x_5$ is at $\infty$. If, in these coordinates, we further gauge-fix so that $x_1 = (1,0,\dots,0)$ and $x_2 = 0$, then we recover the second line of (\ref{averaged}).

\begin{figure}[ht]
\begin{center}
\includegraphics[width=.5\textwidth]{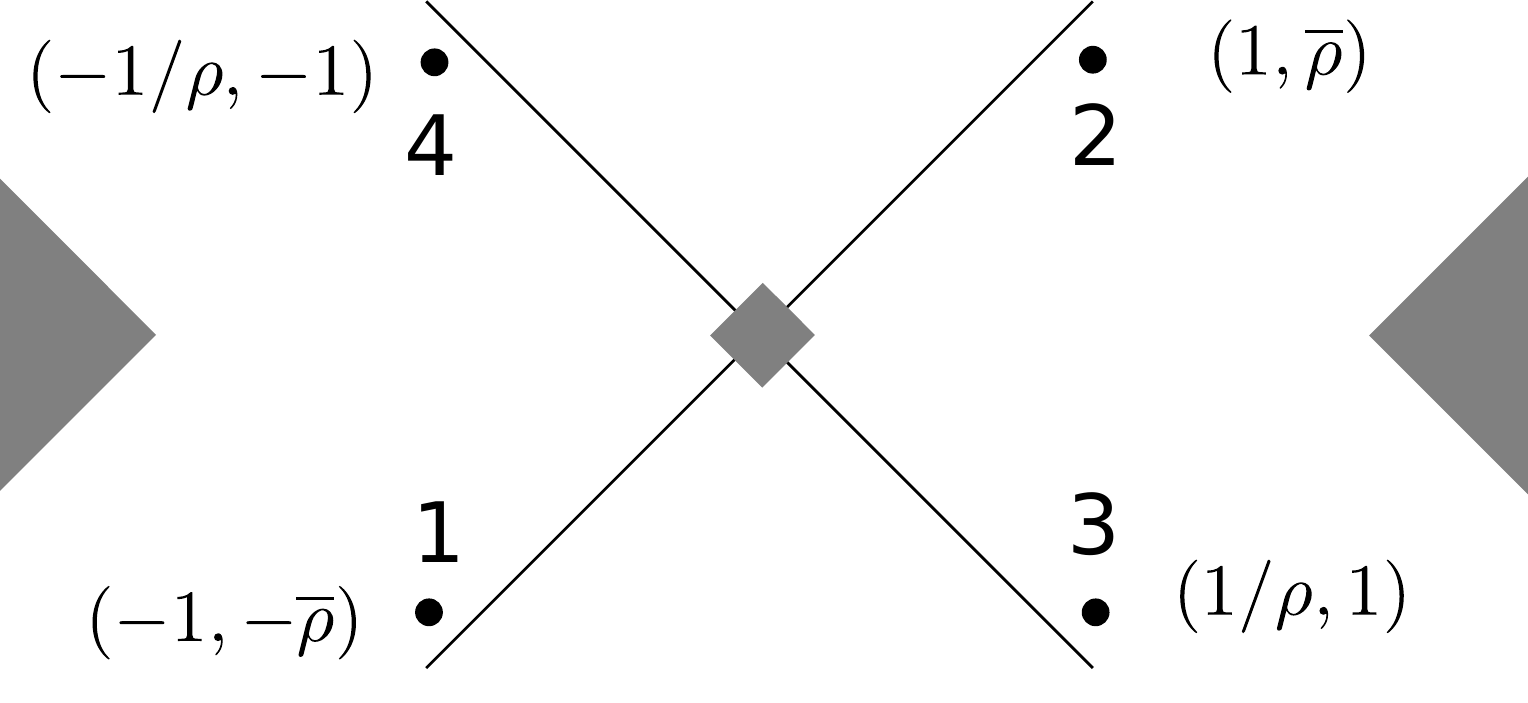}
\caption{\small{The configuration of points that we choose, with $(u,v)$ coordinates indicated. The grey region is spacelike separated from the four points. The 2$d$ slice shown is the plane where the four points are located. As we move the slice outwards in the transverse directions away from this plane, the inner and outer grey regions grow and eventually merge, see figure~\ref{threedintegrationregion}.}}\label{figrhorhobar}
\end{center}
\end{figure}

Instead of picking the gauge that takes us back to (\ref{averaged}), we will pick a different gauge where we fix $x_1\cdots x_4$ to locations determined by the cross ratios, and then integrate over the location of the fifth point, subject to the constraint that it should be spacelike separated from the others. More precisely, we choose the points $x_1\cdots x_4$ to be located in a 2$d$ plane and located as in figure~\ref{figrhorhobar}. The standard conformal cross ratios for this configuration are
\be
\chi = \frac{4\rho}{(1+\rho)^2},\hspace{20pt} \bar{\chi} = \frac{4\bar{\rho}}{(1+\bar{\rho})^2}.
\ee
The advantage of this gauge choice is that we have now cleanly separated the cross ratio degrees of freedom from the other integration variables. The non-cross ratio variables are simply the location of $x_5$.

\subsection{Evaluating the integral for small cross ratios}

We would like to do the integral over $x_5$:
\begin{align}
H_{\Delta,J}(x_i) &\equiv \int_{\mathrm{spacelike}}d^dx_5 \frac{1}{|x_{12}|^{\tl\De_1+\tl\De_2-\tl\De}|x_{15}|^{\tl\De_1+\tl\De-\tl\De_2}|x_{25}|^{\tl\De_2+\tl\De-\tl\De_1}}\notag\\ &\hspace{100pt}\x\frac{B_J(-\eta)}{|x_{34}|^{\tl\De_3+\tl\De_4 - \De}|x_{35}|^{ \tl\De_3+ \De- \tl\De_4}|x_{45}|^{\tl\De_4+ \De -  \tl\De_3}}.\label{Hdef}
\end{align}
Here, we assume that $x_1,\dots, x_4$ are configured as in figure~\ref{figrhorhobar}, and $x_5$ ranges over all points on the cylinder that are spacelike separated from these four points.

By the usual logic of shadow integrals (together with the fact that being spacelike separated from all four other points is a conformally covariant notion), this integral is conformally covariant with weights $\tl \De_1,\dots, \tl \De_4$ for the four external points. Let us strip off some factors with the same external weights to obtain a function of conformal cross ratios $\chi,\b\chi$ alone:
\begin{align}
\label{eq:stripofffactors}
H_{\De,J}(x_i) &= T^{\tl\De_i}(x_i) H_{\De,J}(\chi,\b\chi) = \frac{1}{|x_{12}|^{2d}|x_{34}|^{2d}}\frac{1}{T^{\De_i}(x_i)}H_{\Delta,J}(\chi,\b\chi),
\end{align}
where we distinguish $H_{\Delta,J}(x_i)$ and $H_{\Delta,J}(\chi,\b\chi)$ by their arguments, and
\begin{align}
T^{\De_i}(x_i) &\equiv
\frac{1}{|x_{12}|^{\De_1+\De_2}|x_{34}|^{\De_3+\De_4}}\left(\frac{|x_{14}|}{|x_{24}|}\right)^{\De_2-\De_1}\left(\frac{|x_{14}|}{|x_{13}|}\right)^{\De_3-\De_4}.
\label{eq:defoft}
\end{align}
Note that we take the absolute value of all the intervals $|x_{ij}|=|(x_{ij}^2)|^{1/2}$, even though $x_{14}$ is timelike. This is because $H_{\De,J}(x_i)$ is manifestly real when $\De_i,\De,J$ are real, and we would like $ H_{\De,J}(\chi,\bar{\chi})$ to inherit this property.

\begin{figure}[ht]
\begin{center}
\includegraphics[width=.9\textwidth]{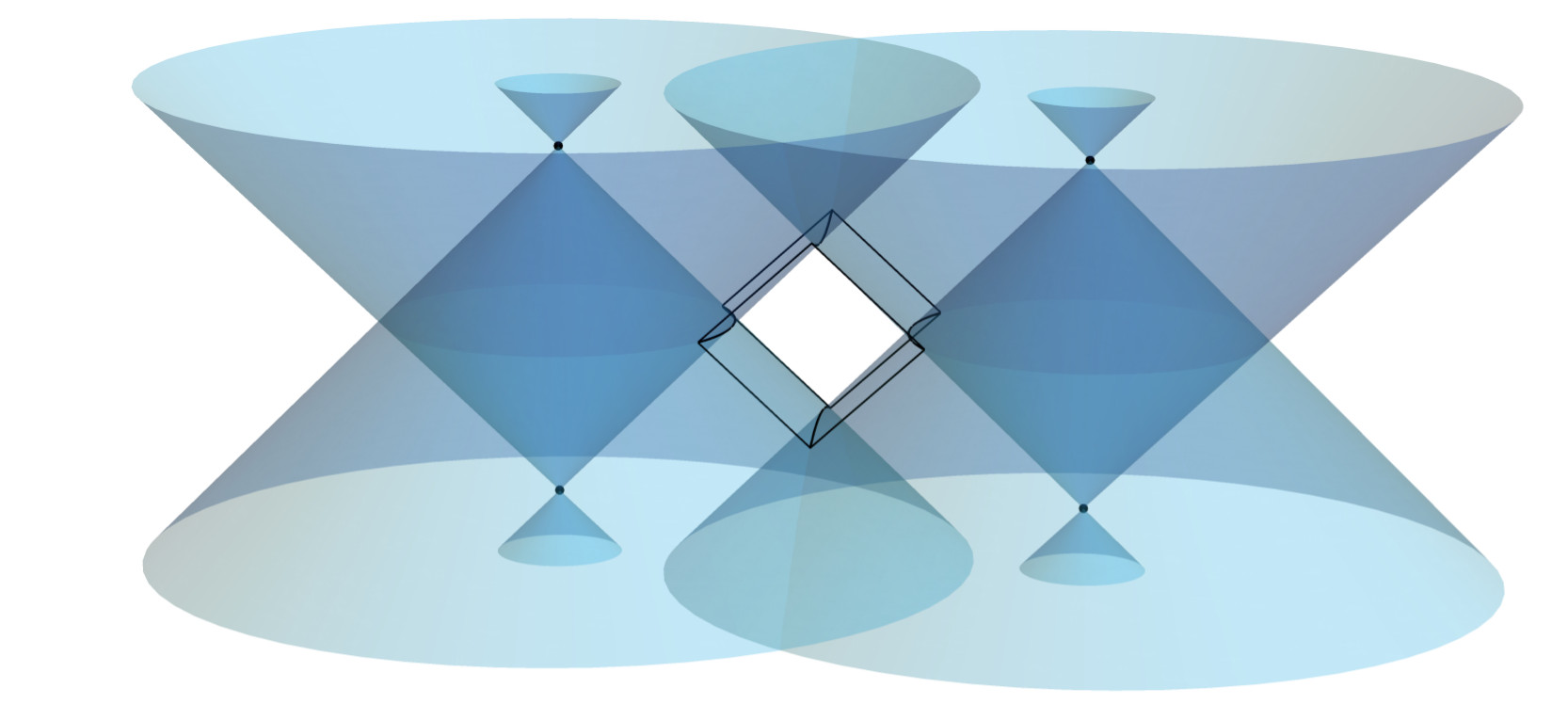}
\caption{\small{The region of integration for $x_5$ is the exterior of the lightcones of the four operators. In the limit of small cross-ratios, the integral is dominated by the region inside the black-outlined box. The width of the box in the transverse directions is large enough to detect the curvature of the lightcones, but small enough not to detect their full geometry.}}\label{threedintegrationregion}
\end{center}
\end{figure}

The integrand in (\ref{Hdef}) is an eigenfunction of the two-particle quadratic and quartic conformal Casimirs (with eigenvalues determined by $\Delta,J$) acting on either $1+2$ or $3+4$. Thus, $H_{\Delta,J}(\chi,\bar{\chi})$ will have the same property. Solutions to these Casimir equations are determined by their behavior for small values of the cross ratios. So we can pin down $H_{\Delta,J}(\chi,\bar{\chi})$ exactly by evaluating it for small $\chi,\b \chi$. In our $\rho,\b \rho$ coordinates, we can reach this regime by taking $\rho\ll 1$ and $\b \rho\gg 1$, so that\footnote{Note that this corresponds to starting with the standard Euclidean configuration described in~\cite{Pappadopulo:2012jk,Hogervorst:2013sma} with $\rho=\bar \rho$, and then applying a large Lorentzian relative boost between the points $1,2$ and $3,4$. This highly boosted configuration of cross-ratios played an important role in the recent causality-based proof of the ANEC~\cite{Hartman:2016lgu}.}
\be
\chi \approx 4\rho, \hspace{20pt} \b \chi \approx \frac{4}{\b \rho}.
\ee
For these small values of the cross ratios, it turns out to be straightforward to evaluate the $x_5$ integral. The integral is dominated by a region where the transverse separation of $x_5$ from the plane of the other four points is small enough that the lightcones of the four external operators can be approximated as simpler shapes. This region is illustrated in figure~\ref{threedintegrationregion}.

We will describe $x_5$ by coordinates $u,v$ in the plane of the other four points, and a radius $r$ in the transverse directions. To organize the integral for small values of the cross ratios, it is helpful to introduce a small parameter $\epsilon \ll 1$, where we take $\rho \sim \epsilon$ and $\b \rho \sim \epsilon^{-1}$ with some fixed product $\rho \b \rho$. The important region of the integral comes from $u,v$ of order one and $r$ of order $1/\sqrt{\epsilon}$. As we will see, in this region one can show that $-\eta$ is large, of order $1/\epsilon$. To summarize, we have
\be
\b\rho,\frac{1}{\rho},-\eta,r^2\sim \frac{1}{\epsilon},\hspace{20pt} u,v \sim 1.
\ee

These scalings allow us to simplify the integral considerably. Since $-\eta$ is large, we can approximate $B_J(-\eta)$ using (\ref{largeyB}). Also, for small $\epsilon$, the quantity $\eta<0$ is determined by a simplified formula that follows from expanding (\ref{defeta}):
\be
4\eta^2 \approx \frac{\bar{\rho}}{\rho}\frac{(\frac{1}{\rho}+r^2)^2}{(\frac{1-v}{\rho}+r^2)(\frac{1+v}{\rho}+r^2)}\frac{(\bar{\rho}+r^2)^2}{(\bar{\rho}(1-u)+r^2)(\bar{\rho}(1+u)+r^2)}.
\ee
The rest of the integrand can also be simplified, by keeping only the terms of order $1/\sqrt\epsilon$ in the distances $|x_{ij}|$. For example, $|x_{15}|\approx (\b\rho(1+u)+r^2)^{1/2}$. After making these approximations, the $u$ and $v$ dependence of the integrand factorizes, as does the region of integration. For example, the $v$ integral is of the form
\be
\int_{-1-\rho r^2}^{1+\rho r^2}dv (1-v+\rho r^2)^{a}(1+v+\rho r^2)^{b} = (2+2\rho r^2)^{1+a+b}\frac{\Gamma(1+a)\Gamma(1+b)}{\Gamma(2+a+b)}.
\ee
Note that in general, the region of integration is not factorized in $u$ and $v$, since the boundaries of the light cones at finite transverse separation are curved in the $u,v$ plane. However, the transverse separations $r\sim 1/\sqrt{\epsilon}$ are small enough that the edges of the ``inner'' region remain straight in the $u,v$ plane, although with a separation that depends on $r$.

After doing the $u,v$ integrals, the final integral over $r$ can be done by changing variables to $y = \rho r^2$ and using
\begin{align}
(\rho\bar{\rho})^{\frac{\tl\Delta-1}{2}}\int_0^\infty dy y^{\frac{d-4}{2}}(\bar{\rho}\rho+y)^{1-\tl\Delta}(1+y)^{1-\Delta} &=\frac{\Gamma(\frac{d}{2}{-}1)^2}{\Gamma(d{-}2)}{}_2F_1\p{\tl\Delta{-}1,\Delta{-}1,\frac{d{-}1}{2},\frac{1{-}x}{2}},
\end{align}
where $x = \frac{1}{2}(\sqrt{\rho\bar{\rho}}+\frac{1}{\sqrt{\rho\bar{\rho}}})$ and as always $\tl\Delta = d-\Delta$.

Collecting factors of $\rho$ and $\b \rho$ and translating to cross ratios, one finds that for $\chi,\b\chi\ll 1$, 
\be
H_{\Delta,J}(\chi,\b\chi)\approx (\text{const.}) \left(\chi\b\chi\right)^{\frac{J+d-1}{2}}{}_2F_1\p{\tl\Delta{-}1,\Delta{-}1,\frac{d{-}1}{2},\frac{1{-}x}{2}},
\ee
where $x = \frac{1}{2}(\sqrt{\chi/\b\chi} + \sqrt{\b\chi/\chi})$. More precisely, we have this behavior plus multiplicative corrections that are analytic at $\chi = \b\chi = 0$. This behavior determines our solution to the Casimir equations. It takes the form similar to that of a standard conformal block, but with ``dimension'' equal to $J+d-1$ and ``spin'' equal to $1-\tl\De = \Delta-d+1$.

It is useful to report the constant of proportionality by giving the behavior for $\chi \ll \b \chi \ll 1$, where the hypergeometric function simplifies. We find
\begin{align}
\label{eq:eqnfora}
H_{\Delta,J}&(\chi,\b\chi) \approx a_{\De,J}\,(\chi \b \chi)^{\frac{J+d-1}{2}}\left(\frac{\chi}{\b\chi}\right)^{-\frac{\Delta-d+1}{2}}\\
a_{\De,J}&\equiv \frac{1}{2}(2\pi)^{d-2}\frac{\Gamma(J+1)}{\Gamma(J+\frac{d}{2})}\frac{\Gamma(\Delta-\frac{d}{2})}{\Gamma(\Delta-1)}\frac{\Gamma(\frac{\De_{12}+J+\De}{2})\Gamma(\frac{\De_{21}+J+\De}{2})\Gamma(\frac{\De_{34}+J+\tl\De}{2})\Gamma(\frac{\De_{43}+J+\tl\De}{2})}{\Gamma(J+\Delta)\Gamma(J+d-\De)},\nn
\end{align}
where $\Delta_{ij} \equiv \Delta_i - \Delta_j$. Comparing to (\ref{eq:blockdoublelimit}), this determines
\begin{align}
\label{eq:equationforhhat}
H_{\De,J}(\chi,\bar \chi) &= a_{\De,J} G^{\tl \De_i}_{J+d-1,\De-d+1}(\chi,\bar\chi).
\end{align}

\subsection{Writing in terms of cross ratios}
As a final step, we can now write the formula for $I_{\Delta,J}$ as an integral over cross ratios only. The result of the last section is that 
\begin{align}
I_{\Delta,J} &\supset-\frac{\hat C_J(1)}{\vol(S^{d-2})}(-1)^J\int_{4>1,2>3} \frac{d^d x_1 \cdots d^d x_4}{\vol(\SO(d,2))}  \<[O_3,O_2][O_1,O_4]\> H_{\De,J}(x_i)
\\
&= -\frac{\hat C_J(1)}{\vol(S^{d-2})}(-1)^J\int_{4>1,2>3} \frac{d^d x_1 \cdots d^d x_4}{\vol(\SO(d,2))} \frac{1}{|x_{12}|^{2d}|x_{34}|^{2d}} \frac{\<[O_3,O_2][O_1,O_4]\>}{T^{\De_i}(x_i)} H_{\De,J}(\chi,\bar\chi),\notag
\end{align}
where $H_{\De,J}$ is the particular solution to the conformal Casimir equation (\ref{eq:equationforhhat}).

Let us now gauge fix this integral in the configuration of figure~\ref{figrhorhobar}. Parameterizing everything in terms of 
 $\chi=\frac{4\rho}{(1+\rho)^2}$ and $\bar\chi=\frac{4\bar\rho}{(1+\bar \rho)^2}$, the gauge-fixed measure becomes
\begin{align}
\int \frac{d^d x_i}{\vol(\SO(d,2))}\frac{1}{|x_{12}|^{2d}|x_{34}|^{2d}} 
\quad
&\to
\quad
 \frac{1}{2\vol(\SO(d{-}2))} \int_0^1 \int_0^1  \frac 1 2 \frac{d\chi d\bar \chi}{(\chi \bar \chi)^d} \left|\frac{\bar \chi-\chi}{2}\right|^{d-2}.
 \label{eq:integrationmeasure}
\end{align}
Let us make some comments about this result. The quantities inside the integral come from a Faddeev-Popov determinant.\footnote{To be consistent with our convention that gauge-fixing three points to $0,1,\oo$ should give determinant $1$, we  must additionally divide by the determinant associated with that gauge fixing.}  The factor $\vol(\SO(d{-}2))$ is the volume of the group of transverse rotations. The extra factor of $\frac 1 2$  is because of an additional discrete symmetry that relates two configurations in our integration range. Specifically, there exists an element of the identity component of $\SO(d,2)$ that exchanges $\rho\leftrightarrow 1/\bar \rho$, or equivalently $\chi\leftrightarrow \b\chi$.  In the plane of the four points, we can achieve this with an inversion followed by a dilatation and boost
\begin{align}
(u,v) &\mapsto \p{\frac {\bar \rho} v, \frac 1 {\rho u}}.
\end{align}
In two dimensions, an inversion is not continuously connected to the identity. However, in higher dimensions, we can accompany it with a reflection in a transverse direction to obtain something continuously connected to the identity. (Different choices of reflection are related by conjugating by the transverse rotation group $\SO(d{-}2)$.) Hence, to to avoid double-counting configurations modulo gauge transformations in (\ref{eq:integrationmeasure}), we must divide by $2$. We could alternatively restrict the integration region to $\chi \leq \bar\chi$ or $\bar\chi \leq \chi$.

We can now write a final expression for $I_{\Delta,J}$, as
\begin{align}\label{finalI}
I_{\Delta,J} = \alpha_{\Delta,J} \Bigg[(-1)^J&\int_0^1\int_0^1 \frac{d\chi d\bar \chi}{(\chi \bar \chi)^d} |\chi-\b\chi|^{d-2} G^{\tl\De_i}_{J+d-1,\Delta-d+1}(\chi,\bar\chi)\frac{\<[O_3,O_2][O_1,O_4]\>}{T^{\De_i}}  \\
&+\int_{-\infty}^0\int_{-\infty}^0\frac{d\chi d\bar \chi}{(\chi \bar \chi)^d} |\chi-\b\chi|^{d-2} \hat G^{\tl\De_i}_{J+d-1,\Delta-d+1}(\chi,\bar\chi)\frac{\<[O_4,O_2][O_1,O_3]\>}{T^{\De_i}} \Bigg].\notag
\end{align}
In writing this expression we have also added back in the contribution from the second line of (\ref{averaged}). The function $\hat G_{\Delta,J}$ is defined as the conformal block normalized so that for negative cross ratios satisfying $|\chi| \ll |\b \chi| \ll 1$ we have the behavior $(-\chi)^{\frac{\Delta-J}{2}}(-\b\chi)^{\frac{\Delta+J}{2}}$.  Note that this differs by a phase from the continuation of $G_{\Delta,J}$ to negative values of $\chi,\b\chi$. The constant out front is
\be\label{eq:eqnforalpha}
\alpha_{\Delta,J} = -\frac{a_{\Delta,J}}{2^{d}}\frac{\hat C_J(1)}{\vol(\SO(d{-}1))},
\ee
where $a_{\Delta,J}$ is defined in (\ref{eq:eqnfora}). In order to compare to Caron-Huot, we should use (\ref{translate}) to convert from the inner product $I_{\Delta,J}$ between the four-point function and a partial wave to the coefficient $c(J,\Delta)$ in the partial wave expansion of the four-point function. The equation for $c(J,\Delta)$ is then the same as (\ref{finalI}), but with the constant out front replaced by
\be\label{alphadef}
\frac{\alpha_{\Delta,J}}{n_{\Delta,J}}K^{\Delta_3,\Delta_4}_{\tl\De,J} = -(-1)^{J}\frac{\Gamma(\frac{J+\Delta+\Delta_{12}}{2})\Gamma(\frac{J+\Delta-\Delta_{12}}{2})\Gamma(\frac{J+\Delta+\Delta_{34}}{2})\Gamma(\frac{J+\Delta-\Delta_{34}}{2})}{16\pi^2\Gamma(J+\Delta-1)\Gamma(J+\Delta)}.
\ee
We can relate the double commutator to Caron-Huot's ``double discontinuity'' dDisc by defining a stripped four-point function as
\begin{align}
\<O_1 O_2 O_3 O_4\> &= \frac{1}{(x_{12}^2)^{\frac{\De_1+\De_2}{2}}(x_{34}^2)^{\frac{\De_3+\De_4}{2}}}\left(\frac{x_{14}^2}{x_{24}^2}\right)^{\frac{\De_2-\De_1}{2}}\left(\frac{x_{14}^2}{x_{13}^2}\right)^{\frac{\De_3-\De_4}{2}} g(\chi,\bar \chi).
\end{align}
Applying the appropriate $i\e$ prescriptions in the configuration of figure~\ref{figrhorhobar}, we find
\begin{align}
\frac{\<[O_3,O_2][O_1,O_4]\>}{T^{\De_i}}
&= -2\cos(\pi\tfrac{\De_2-\De_1+\De_3-\De_4}{2})\, g(\chi,\bar\chi) + e^{i\pi\tfrac{\De_2-\De_1+\De_3-\De_4}{2}} g^\circlearrowleft(\chi,\bar\chi) \nn\\
&\qquad+ e^{-i\pi\tfrac{\De_2-\De_1+\De_3-\De_4}{2}}g^\circlearrowright(\chi,\bar\chi)\nn\\
&\equiv -2\,\mathrm{dDisc}[g(\chi,\bar\chi)],
\end{align}
where $g^{\circlearrowleft}$ or $g^{\circlearrowright}$ indicates we should take $\bar\chi$ around $1$ in the direction shown, leaving $\chi$ held fixed. Note that the minus sign in this formula is because our convention for operator ordering is the one natural for the standard quantization of the theory in a global Minkowski time, not Rindler time. Similarly,
\begin{align}
\frac{\<[O_4,O_2][O_1,O_3]\>}{T^{\De_i}} &= -2\cos\p{\pi\tfrac{\Delta_2-\Delta_1+\Delta_4-\Delta_3}{2}}g(\chi,\b\chi) + e^{i\pi\frac{\Delta_3-\Delta_4+\Delta_2-\Delta_1}{2}}g^\circlearrowright(\chi,\b\chi)\notag\\&\qquad + e^{-i\pi\frac{\Delta_3-\Delta_4+\Delta_2-\Delta_1}{2}}g^\circlearrowleft(\chi,\b\chi),
\end{align}
where now $g^\circlearrowleft$ or $g^\circlearrowright$ indicates we should take $\b\chi$ around $-\infty$ in the direction shown, leaving $\chi$ held fixed.

Finally, note that the formula in~\cite{Caron-Huot:2017vep} contains the block $G_{J+d-1,\De-d+1}^{\De_i}(\chi,\bar\chi)$ with un-tilded external dimensions $\De_i$, and also some additional factors in the measure. In our formula, these come from the identity~\cite{Dolan:2011dv}
\begin{align}\label{fromdolan}
G_{J+d-1,\De-d+1}^{\tl \De_i}(\chi,\bar\chi) &= ((1-\chi)(1-\bar \chi))^{\frac{\De_2-\De_1+\De_3-\De_4}{2}}G_{J+d-1,\De-d+1}^{\De_i}(\chi).
\end{align}
With this understanding, and using (\ref{alphadef}), we find that (\ref{finalI}) agrees precisely with the formula in~\cite{Caron-Huot:2017vep}.\footnote{We have reversed the subscripts on $G_{\De,J}$ relative to~\cite{Caron-Huot:2017vep}.}

\section{One dimension}
There is a one-dimensional analog of Caron-Huot's formula, although it is less powerful than in higher dimensions. In one dimension, the complete set of partial waves includes a discrete series, in addition to the principal continuous series. All wave functions are related by analytic continuation in $\Delta$, so the same function $I_{\Delta}$ describes the inner product of both principal series and discrete series states with the four-point correlator. This function has poles for positive $\text{Re}(\Delta)$ that correspond to physical operators of the theory.

The formula we can derive is for a different function $\tl{I}_\Delta$ that agrees with $I_{\Delta}$ for the discrete series of integer $\Delta$, but has the additional property that it is analytic (without poles) for $\text{Re}(\Delta)>1$. These properties seems somewhat arbitrary, but there is a good reason for the existence of such a function. As described in~\cite{Maldacena:2016hyu}, when one continues to the Regge limit in a one dimensional $\SL(2,\R)$ invariant theory, the discrete states give growing contributions that naively form a divergent series. If, before continuing to the Regge limit, we write this sum as an integral over a contour that consists of small circles around the discrete states at positive integer $\Delta$, and if we use $\tl{I}_\Delta$ instead of $I_\Delta$ in this expression, then we can pull the contour to the left towards a region with bounded Regge behavior. The absence of poles in $\tl{I}_\Delta$ allows us to do this continuation without picking up growing contributions that would spoil boundedness in the Regge limit.

In one dimension we do not have light-cone coordinates. However, because the conformal blocks are simple, it is easy enough to derive the formula directly in cross ratio space.\footnote{We assume time-reversal
symmetry, so that the four-point function is a function of the cross ratio only.  Without time-reversal symmetry, there is an additional discrete invariant.  For the shadow representation without the assumption of
time-reversal symmetry, see~\cite{KB}.} In one dimension the global conformal group is $\SL(2,\R)$, and a four-point function depends on a single cross ratio. The wave functions are given by the shadow representation\footnote{As
in our $d=2$ discussion in section~\ref{twod}, and as in the usual SYK model, we will slightly simplify this discussion by assuming that the four external operators have the same dimension $\Delta_O$.}
\begin{align}
\Psi^{\Delta_O}_{\Delta,J}(\tau_i) &= \frac{1}{|\tau_{12}|^{2\Delta_O}|\tau_{34}|^{2\Delta_O}}\Psi_{\Delta,J}(\chi)\notag\\
\Psi_{\Delta,0}(\chi) &= \int d\tau_5 \left(\frac{|\tau_{12}|}{|\tau_{15}||\tau_{25}|}\right)^\Delta\left(\frac{|\tau_{34}|}{|\tau_{35}||\tau_{45}|}\right)^{1-\Delta}\\
\Psi_{\Delta,1}(\chi) &= \int d\tau_5 \left(\frac{|\tau_{12}|}{|\tau_{15}||\tau_{25}|}\right)^\Delta\left(\frac{|\tau_{34}|}{|\tau_{35}||\tau_{45}|}\right)^{1-\Delta}\text{sgn}(\tau_{12}\tau_{15}\tau_{25}\tau_{34}\tau_{35}\tau_{45}).\notag
\end{align}
The functions $\Psi_{\Delta,J}$ on the second two lines depend only on the cross ratio $\chi = \frac{\tau_{12}\tau_{34}}{\tau_{13}\tau_{24}}$. The ``spin'' $J$ takes two possible values, 0 and 1, {and to get a complete basis of functions of $\chi$, we have to consider both. The functions with $J = 0$ are symmetric under the transformation $\chi \rightarrow \chi/(\chi-1)$, and the functions with $J = 1$ are antisymmetric. The reason that we refer to $J$ as spin is that these expressions are actually the analytic continuation in dimension from the higher dimensional shadow integrals (\ref{shadowgend}). In one dimension $\eta$ reduces to the product of sgn factors in $\Psi_{\Delta,1}$, so we can understand this factor as $\hat C_1(\eta) = \eta$. The fact that we don't have other functions is also consistent with the higher dimensional formulas, since when $d = 1$ and we consider a value of $J\ge 2$, we have $\hat C_J(\pm 1) = 0$.}

The complete set of partial waves corresponds to the principal series $\Delta = \frac{1}{2}+ir$ in addition to the even positive integers for $\Psi_{\Delta,0}$ and the odd positive integers for $\Psi_{\Delta,1}$. We will be concerned only with the discrete series here. By evaluating the shadow integrals for integer $\Delta$, one finds a uniform expression for these discrete states as\footnote{This form of $\Psi_n(\chi)$ for $\chi<1$
was obtained in 
\cite{Maldacena:2016hyu}.  (It arises  from eqns.~(3.65), (3.66) of that paper by setting $h$ equal to an integer $n$.) The result for $\chi>1$ then follows from the matching conditions  described in~\cite{Maldacena:2016hyu} between the $\chi<1$ and $\chi>1$ wavefunctions.  Those wavefunctions
behave near $\chi=1$ as $a+b\log |1-\chi|$, and the matching conditions say that the coefficients $a$ and $b$ are equal on the two sides.  The following derivation will depend only on the fact that the same
function $k_{2n}(\chi)$ appears in both lines of eqns.~(\ref{d3}), not on any further properties of this function.} 
\begin{align}
\Psi_n(\chi) &= 2\frac{\Gamma(n)^2}{\Gamma(2n)} k_{2n}(\chi),  &-\infty < \chi < 1, \label{d3} \nn\\
\Psi_n(\chi) &= \frac{\Gamma(n)^2}{\Gamma(2n)} \left[k_{2n}(\chi+i\epsilon) + k_{2n}(\chi-i\epsilon)\right], &1< \chi < \infty.
\end{align}
In this section we will continue to use the notation for the $\SL(2,\R)$ block
\begin{align}
k_{2h}(\chi) \equiv \chi^h {}_2F_1(h,h,2h,\chi),\quad \hat{k}_{2h}(\chi) \equiv (-\chi)^h {}_2F_1(h,h,2h,\chi).
\end{align}
The notation $\Psi_n$ is defined for positive integer $n = 1,2,\dots$. For even integer $n$ it is equal to $\Psi_{n,0}$ and for odd integer $n$ it is equal to $\Psi_{n,1}$.

The inner product of these wave functions with the four-point function is defined as
\be
I_{n} = \int_{-\infty}^\infty \frac{d\chi}{\chi^{2}} g(\chi)\Psi_{n}(\chi), \hspace{40pt}g(\chi) \equiv \frac{\<O_1O_2O_3O_4\>}{\< O_1O_2\> \< O_3O_4\>},
\ee
where $g(\chi)$ is the stripped four-point function. It is piecewise analytic, consisting of three different analytic functions in the regions $-\infty < \chi<0$, $0<\chi<1$ and $1<\chi<\infty$. In the region $1<\chi<\infty$, we insert (\ref{d3}) and then deform the two terms in opposite half-planes to the region $-\infty <\chi <1$. We then find
\be
I_n = \frac{\Gamma(n)^2}{\Gamma(2n)}\int_{-\infty}^1 \frac{d\chi}{\chi^{2}} k_{2n}(\chi)\text{dDisc}[g(\chi)].
\ee
Here $\text{dDisc}(\chi)$ is defined as
\be
\text{dDisc}[g(\chi)] = 2g(\chi) - g^{\curvearrowleft}(\chi)-g^{\text{\rotatebox[origin=c]{180}{\reflectbox{$\curvearrowleft$}}}}(\chi),
\ee 
where $g^{\curvearrowleft}$ and $g^{\text{\rotatebox[origin=c]{180}{\reflectbox{$\curvearrowleft$}}}}$ are defined by starting with $g(\chi)$ in the region $\chi>1$ and continuing either below or above the real axis to the final value of $\chi <1$.

So far our manipulations are valid for integer $n$, but we can now continue in $n$. We have to take care with defining the continuation of $\chi^n$ for negative $\chi$. To define this we first write it as $(-\chi)^n (-1)^n$. Noting that $n$ is even for the discrete states corresponding to $J = 0$ and odd for the states corresponding to $J = 1$, we can write the sign factor as $(-1)^J$. Then
\begin{align}
\tl I_{\Delta,J} &= \frac{\Gamma(n)^2}{\Gamma(2n)}\left[(-1)^J\int_{-\infty}^0 \frac{d\chi}{\chi^{2}}\hat k_{2\Delta}(\chi) \text{dDisc}[g(\chi)]  + \int_{0}^1 \frac{d\chi}{\chi^{2}} k_{2\Delta}(\chi)\text{dDisc}[g(\chi)]\right].
\end{align}
To summarize, we are making two claims about this function. First, it is analytic in $\Delta$ without poles for real part of $\Delta >1$. This is obvious because boundedness in the Regge limit implies that $\text{dDisc}[g(\chi)]$ is bounded by a constant for small $\chi$, and then $\text{Re}(\Delta)>1$ is enough to ensure that the integral converges. Second, for even integer values of $\Delta$ (for $J = 0$) and odd integer values of $\Delta$ (for $J = 1$), this agrees with $I_{\Delta,J}$. This was the content of the above argument.

\section{Discussion}

CFT four-point functions are bounded in the Regge limit~\cite{Maldacena:2015waa}. Just as in the case of amplitudes, nice Regge behavior requires a delicate balance between partial waves. Indeed, an individual conformal block with spin $J$ grows like $e^{(J-1)t}$ in the Regge limit, where $t$ is a boost parameter~\cite{Cornalba:2006xm}. Thus, if we modified the coefficient of a single block with spin $J>1$, we would completely destroy boundedness in the Regge limit. Caron-Huot's formula captures the delicate balance between partial waves by showing that for $J>1$ they fit together into an analytic function of spin with nice properties. This justifies the methods of ``conformal Regge theory"~\cite{Cornalba:2007fs,Costa:2012cb}. It also removes the ambiguities associated with asymptotic series in large-spin perturbation theory~\cite{Komargodski:2012ek,Fitzpatrick:2012yx,Alday:2015eya,Alday:2015ota,Alday:2015ewa,Kaviraj:2015cxa,Alday:2016njk,Simmons-Duffin:2016wlq,Dey:2017fab}, leading to a finite expansion with no need for resummation.\footnote{To compute the large-spin expansion, one applies the Lorentzian inversion formula to the four-point function in an expansion around the double-lightcone limit $\chi, 1-\bar\chi \ll 1$. Reaching the double-lightcone limit from an OPE channel requires summing infinite families of conformal blocks with bounded twist, using e.g.\  the techniques of~\cite{Alday:2016njk,Simmons-Duffin:2016wlq}.} Positivity of the double-commutator in the Lorentzian inversion formula also makes it easy to prove bounds on CFT data like Nachtmann's theorem~\cite{Nachtmann:1973mr,Komargodski:2012ek,Costa:2017twz}.

In this work, we have given a new derivation of Caron-Huot's formula. An advantage of our approach is that analyticity in spin is almost immediate. After performing the contour deformation in (\ref{afterdeform}), it is clear that $I_{\De,J}$ is an analytic function of spin. In addition, the derivation in~\cite{Caron-Huot:2017vep} relied on a surprising identity between analytic continuations of conformal blocks, which we have essentially proved using explicit integral representations for the blocks.

Although we have not focused on this perspective, our inspiration came from thinking about the Regge limit in the SYK model. There, a special relationship between the standard kernel (which is essentially $k_{\Delta,J}$ discussed in section~\ref{shortcut} for the case of mean field theory) and a ``retarded kernel'' (first used in \cite{kitaevfirsttalk}) made it possible to analyze the Regge limit~\cite{Maldacena:2016hyu,Murugan:2017eto}. (A special case of the computation in section~\ref{twod} of the present
paper can be found in Appendix D of~\cite{Murugan:2017eto}.)  What we have done in this paper is to show that this relationship holds in general conformal field theories, not just mean field theory. To make this slightly more explicit, one views the full four-point function of the CFT as a kernel similar to the ladder kernel in SYK\@. The quantity $I_{\Delta,J}$ is related to the eigenvalues of this kernel, as discussed in section~\ref{shortcut}. The analog of the retarded kernel from SYK is essentially the double commutator. The fact that the eigenvalues of these kernels are the same is the content of this paper.

We hope that our derivation points the way to generalizations for external operators with spin and perhaps higher-point functions. A method for deriving Lorentzian inversion formulas for correlators with external spins was recently given in~\cite{Karateev:2017jgd}. The main idea is to integrate by parts with conformally-covariant differential operators to reduce the inversion formula to the scalar case. However, it should be possible to derive a more direct formula, perhaps by combining our derivation with the methods of~\cite{Kravchuk:2017dzd}. A spinning Lorentzian inversion formula would be helpful, for example, for studying correlators of stress-tensors.\footnote{Particularly in holographic theories where the double-commutator kills the contribution of $t$-channel double-trace states at orders $1/N^0$ and $1/N^2$.}

The simplest operators to describe in large-spin perturbation theory are ``double-twist" families~\cite{Komargodski:2012ek,Fitzpatrick:2012yx}.\footnote{``Double-twist" is equivalent to double-trace in large-$N$ theories.} An inversion formula for higher-point functions could make it easier to study multi-twist operators. It is also interesting to ask whether CFT data can be extended to analytic functions of other Dynkin indices of $\SO(d)$ besides $J$, which become important in higher-point functions.

\section*{Acknowledgements}

We thank Simon Caron-Huot, Abhijit Gadde, Denis Karateev, Petr Kravchuk, Eric Perlmutter, and Shu-Heng Shao for discussions. DS is supported by Simons Foundation
grant 385600. DSD is supported by Simons Foundation grant 488657.  EW is supported in part by NSF Grant PHY-1606531.

\appendix

\section{Details on the partial waves}\label{app:partialwaves}
\subsection{Relation to conformal blocks}
The partial wave $\Psi_{\De,J}^{\De_i}(x_i)$ is a sum of two solutions to the conformal Casimir equation with simple behavior near $x_{12}=0$: a conformal block $G_{\De,J}^{\De_i}(x)$ and a shadow block $G_{\tl \De,J}^{\De_i}(x)$~\cite{Dobrev:1977qv,Dolan:2000ut}. For completeness, we will describe the exact expression and also compute the normalization factor $n_{\De,J}$. In the OPE limit $x_{12}\to 0$, the first term $G_{\De,J}^{\De_i}(x)$ behaves how we would expect from naively taking $x_{12}\to 0$ in (\ref{shadow}) inside the integrand. In this limit, $G_{\De,J}^{\De_i}(x)$ comes from the regime of the integral where $x_5$ does not probe the neighborhood near $x_{1,2}$, so that the OPE is valid. To compute its coefficient, we can take $x_{12}\to 0$ in the integrand first and then perform the resulting integral. (Similarly, to compute the second term, we can take $x_{34}\to 0$ before integrating.)

Expanding in small $x_{12}$, we have
\begin{align}
\<O_1(x_1) O_2(x_2) O^{\mu_1 \cdots \mu_J}(x_5)\> 
&\sim |x_{12}|^{\De-\De_1-\De_2-J} x_{12}^{\nu_1}\cdots x_{12}^{\nu_J}\<O_{\nu_1\cdots\nu_J}(x_1)O^{\mu_1\cdots\mu_J}(x_5)\>,
\end{align}
where
\begin{align}
\<O_{\nu_1\cdots\nu_J}(x_1)O^{\mu_1\cdots\mu_J}(x_5)\> &= \frac{I_{(\nu_1}^{\mu_1}(x_{15})\cdots I_{\nu_J)}^{\mu_J}(x_{15})-\textrm{traces}}{|x_{15}|^{2\De}}, \nn\\
I_\nu^\mu(x) &= \de_\nu^\mu - \frac{2x_\nu x^\mu}{x^2}.
\end{align}
Applying this result inside the integrand (\ref{shadow}), we find
\begin{align}
\Psi_{\De,J}^{\De_i}(x_i) &\supset |x_{12}|^{\De-\De_1-\De_2-J} x_{12}^{\nu_1}\cdots x_{12}^{\nu_J} \int d^d x_5 \<O_{\nu_1\cdots\nu_J}(x_1)O^{\mu_1\cdots\mu_J}(x_5)\>\< \tl O_{\mu_1\cdots\mu_J}(x_5) O_3 O_4\> +\dots \nn\\
&= S^{\De_3,\De_4}_{\tl\De,J} |x_{12}|^{\De-\De_1-\De_2-J} x_{12}^{\nu_1}\cdots x_{12}^{\nu_J}\<O_{\nu_1\cdots\nu_J}(x_1) O_3 O_4\> + \dots.
\end{align}
Here, ``$\dots$" represents subleading terms in the $x_{12}\to 0$ limit, and ``$\supset$" means that we are studying one of the two terms in $\Psi_{\De,J}^{\De_i}(x_i)$. The integral on the first line takes the form of a ``shadow transform," where we integrate a two-point function against a three-point function. By conformal invariance, such an integral must be proportional to a three-point function,
\begin{align}
\int dy \<\tl O^{\nu_1\cdots\nu_J}(x) \tl O^{\mu_1\cdots \mu_J}(y)\>\<O_{\mu_1\cdots \mu_J}(y) O_1 O_2\> &= S_{\De,J}^{\De_1,\De_2} \<\tl O^{\nu_1\cdots\nu_J}(x) O_1 O_2\>,
\end{align}
where~\cite{Dobrev:1977qv,Dolan:2000ut}\footnote{The shadow coefficients $S_O^{O_1 O_2}$ are simple to compute using ``weight-shifting operators"~\cite{Karateev:2017jgd,FutureKKSD}.}
\begin{align}
S_{\De,J}^{\De_1,\De_2} &= \frac{\pi^{\frac d 2} \G(\De-\frac d 2)\G(\De+J-1)\G(\frac{\tl\De+\De_1-\De_2+J}{2})\G(\frac{\tl\De+\De_2-\De_1+J}{2})}{\G(\De-1) \G(d-\De+J) \G(\frac{\De+\De_1-\De_2+J}{2})\G(\frac{\De+\De_2-\De_1+J}{2})}.
\label{eq:shadowcoefficient}
\end{align}
So we conclude that the partial wave includes the term
\begin{align}
\label{eq:kcoefficient}
\Psi_{\De,J}^{\De_i}(x_i) &\supset K^{\De_3,\De_4}_{\tl \De,J} G_{\De,J}^{\De_i}(x_i),\quad \textrm{ where }\quad
K^{\De_3,\De_4}_{\tl\De,J} \equiv (-\tfrac 1 2)^J S^{\De_3,\De_4}_{\tl\De,J},
\end{align}
and we have chosen to normalize the conformal block so that
\begin{align}
G_{\De,J}^{\De_i}(0,x,e,\oo) &\sim (-2)^J |x|^{\De-\De_1-\De_2-J} x_{12}^{\nu_1}\cdots x_{12}^{\nu_J} \<O_{\nu_1\cdots\nu_J}(0) O_3(e) O_4(\oo)\> + \dots \nn\\
&= |x|^{\De-\De_1-\De_2} 2^J\hat C_J\p{\frac{x\.e}{|x|}} + \dots,
\end{align}
where $e$ is a unit vector. Performing a similar analysis for $x_{34}\to 0$, to obtain the coefficient of the shadow block, we find the final expression
\begin{align}
\Psi_{\De,J}^{\De_i}(x_i) &= K^{\De_3,\De_4}_{\tl\De,J} G_{\De,J}^{\De_i}(x_i) + K^{\De_1,\De_2}_{\De,J} G_{\tl \De,J}^{\De_i}(x_i).
\end{align}
We use this expression in several places in this paper. A useful fact that follows from this expression is that
\be\label{relationdeltad}
\Psi^{\De_i}_{\tl\De,J} = \frac{K^{\De_3,\De_4}_{\De,J}}{K^{\De_1,\De_2}_{\De,J}}\Psi^{\De_i}_{\De,J}.
\ee

It is conventional to define functions of cross-ratios $\chi,\bar\chi$ alone by stripping off some factors with the same scaling weights as the operators $O_1,\dots,O_4$,
\begin{align}
G^{\De_i}_{\De,J}(x_i) &= \frac{1}{(x_{12}^2)^{\frac{\De_1+\De_2}{2}}(x_{34}^2)^{\frac{\De_3+\De_4}{2}}}\left(\frac{x_{14}^2}{x_{24}^2}\right)^{\frac{\De_2-\De_1}{2}}\left(\frac{x_{14}^2}{x_{13}^2}\right)^{\frac{\De_3-\De_4}{2}} G^{\De_i}_{\De,J}(\chi,\bar \chi).
\end{align}
(The function of cross-ratios $G^{\De_i}_{\De,J}(\chi,\bar \chi)$ actually only depends on the differences $\De_1-\De_2$ and $\De_3-\De_4$.)
The limit $\chi \ll \b\chi \ll 1$ is of particular interest. In this limit, the function of cross ratios becomes
\begin{align}
\label{eq:blockdoublelimit}
G^{\De_i}_{\De,J}(\chi,\bar \chi) &\sim (\chi \bar \chi)^{\frac \De 2} \p{\frac\chi {\b\chi}}^{-\frac J 2},
\qquad
(\chi \ll \bar\chi \ll 1).
\end{align}

\subsection{Normalization}
Finally, let us determine the normalization factor $n_{\De,J}$. This computation was done in appendix A of~\cite{Caron-Huot:2017vep}, but we include it here for completeness. Consider the inner product
\begin{align}
(\Psi^{\De_i}_{\De,J},\Psi^{\tl \De_i}_{\tl \De',J'}) &= \int \frac{d^dx_1\cdots d^d x_4}{\vol(\SO(d{+}1,1))} \Psi_{\De,J}^{\De_i}(x_i) \Psi_{\tl\De',J'}^{\tl\De_i}(x_i) \nn\\
&= \int \frac{d^d x}{\vol(\SO(d{-}1))} \Psi_{\De,J}^{\De_i}(0,x,e,\oo)\Psi_{\tl\De',J'}^{\tl\De_i}(0,x,e,\oo).
\end{align}
where $\De=\frac d 2 + is$ and $\De'=\frac d 2 + is'$ with $s,s'\geq 0$. The result should be proportional to $\de(s-s')$, which can only come from a singularity near $x=0$. One such singularity comes from the term
\begin{align}
(\Psi^{\De_i}_{\De,J},\Psi^{\tl \De_i}_{\tl \De',J'}) &\supset \frac{K^{\De_3,\De_4}_{\tl\De,J}K^{\tl \De_3, \tl \De_4}_{\De',J'}}{\vol(\SO(d{-}1))}\int d^d x G_{\De,J}^{\De_i}(0,x,e,\oo) G_{\tl \De',J'}^{\tl\De_i}(0,x,e,\oo) \nn\\
&= \frac{K^{\De_3,\De_4}_{\tl\De,J}K^{\tl \De_3, \tl \De_4}_{\De',J'}2^{2J}}{\vol(\SO(d{-}1))} \int d^d x |x|^{\De-\De'-d} \hat C_J\p{\frac{x\.e}{|x|}} \hat C_{J'}\p{\frac{x\.e}{|x|}} + \dots
\nn\\
&=\frac{K^{\De_3,\De_4}_{\tl\De,J}K^{\tl \De_3, \tl \De_4}_{\De,J}\vol(S^{d-2})}{\vol(\SO(d{-}1))} \frac{(2J+d-2) \pi \G(J+1)\G(J+d-2)}{2^{d-2}\G(J+\frac d 2)^2}  \pi \de(s-s') \de_{JJ'} + \dots
\label{eq:normcalculation}
\nn\\
\end{align}
where ``$\dots$" represents nonsingular contributions away from $x=0$ that must drop out in the final result. The product
\begin{align}
K_{\tl \De,J}^{\De_3,\De_4} K_{\De,J}^{\tl \De_3 \tl \De_4}
= \frac{1}{2^{2J}}\frac{\pi^d \G(\De-\frac d 2)\G(\tl \De-\frac d 2)}{(\De+J-1)(\tl \De+J-1)\G(\De-1)\G(\tl \De-1)}.
\end{align}
is independent of $\De_3$ and $\De_4$.  An equal contribution comes from the term $G^{\De_i}_{\tl \De,J} G^{\tl \De_i}_{\De',J}$, giving an additional factor of 2. Overall, we have
\begin{align}\label{normcoeff}
n_{\De,J} &= \frac{K_{\tl \De,J}^{\De_3,\De_4} K_{\De,J}^{\tl \De_3 \tl \De_4}\vol(S^{d-2})}{\vol(\SO(d{-}1))} \frac{(2J+d-2) \pi \G(J+1)\G(J+d-2)}{2^{d-2}\G(J+\frac d 2)^2}.
\end{align}

\subsection{Completeness}\label{onethinghere}
In this section we will discuss the completeness of the partial waves. A first step is to describe the inner product, since this defines orthogonality and also establishes the Hilbert space of square-integrable functions in which we are trying to prove completeness. In the main text of the paper we did not discuss an inner product exactly, but we did discuss a closely related bilinear pairing (\ref{innerproddef}), which  after gauge-fixing to cross-ratios 
(as will be convenient in this appendix) reduces to
\be
\left(\Psi^{\tl\De_i}_{\tl\De',J'},\Psi^{\De_i}_{\De,J}\right) = \frac{1}{2\vol(\SO(d{-}2))}\int \frac{d^2\chi}{|\chi|^{2d}}|\text{Im}(\chi)|^{d-2}\Psi^{\tl\De_i}_{\tl\De',J'}(\chi,\b\chi)\Psi^{\De_i}_{\De,J}(\chi,\b\chi).
\ee 
(The factor of $1/2$ is because we are letting $\Im(\chi)$ be both positive and negative.)
We would like to interpret this pairing as an inner product. 

\subsubsection*{External dimensions in the principal series}
We will start by discussing the unphysical case where the external dimensions are in the principal series, $\Delta_i = \frac{d}{2}+i r_i$. We will come back to the physical case of real external dimensions below. If the internal dimension $\Delta$ is also in the principal series, then the above is actually a complex inner product
\begin{align}\label{innerprodone}
\left(\Psi^{\tl\De_i}_{\tl\De',J'},\Psi^{\De_i}_{\De,J}\right) &= \left\langle \Psi^{\De_i}_{\De',J'},\Psi^{\De_i}_{\De,J}\right\rangle,\nn\\
\langle F,G\rangle &\equiv \frac{1}{2\vol(\SO(d{-}2))}\int \frac{d^2\chi}{|\chi|^{2d}}|\text{Im}(\chi)|^{d-2}\overline{F}(\chi,\b\chi)G(\chi,\b\chi).
\end{align}
Here we are simply using that if a dimension $\Delta$ is in the principal series, then $d-\Delta$ is the same thing as the complex conjugate of $\Delta$, i.e.\ $\tl\Delta = \overline{\Delta}$.

We will now argue that the partial waves with integer $J$ and internal dimension $\Delta$ in the principal series with $r>0$ are complete for the Hilbert space defined by this inner product, and with the restriction of symmetry under $\chi\leftrightarrow \b\chi$. The argument is based on the idea that the normalizable eigenfunctions of commuting Hermitian operators should be complete. In our case we can consider the operators to be the quadratic and quartic Casimir differential operators. These operators are Hermitian with respect to the inner product (\ref{innerprodone}). For fixed eigenvalues of the two Casimirs, there are eight linearly independent solutions. The requirement that the functions be single valued around $\chi = 0$ and $\chi = 1$ and symmetric under $\chi\leftrightarrow \b\chi$ reduces us to a single solution, which is the partial wave $\Psi_{\Delta,J}$, with $\Delta,J$ related to the eigenvalues of the two Casimirs, and $J$ constrained to be an integer.

Finding a complete set of functions then reduces to the problem of finding the full set of values $\Delta,J$ such that the corresponding partial wave is square-integrable. When the external dimensions are in the principal series, the only constraint comes from imposing normalizability at $\chi =0$. In order for a function to be (continuum) normalizable with respect to (\ref{innerprodone}), it must vanish at least as fast as $|\chi|^{d/2}$. Now, for small $|\chi|$, the partial waves have two terms with the behavior (ignoring the angular dependence)
\be
\Psi_{\De,J}^{\Delta_i} \sim K^{\De_3,\De_4}_{\tl \De,J} |\chi|^{\Delta} + K^{\De_1,\De_2}_{\De,J} |\chi|^{d-\Delta}.
\ee
In order for both of these to be continuum normalizable, we need $\Delta = \frac{d}{2}+ir$ for some real $r$. It follows from (\ref{relationdeltad}) that the partial waves with $r<0$ are proportional to the partial waves with $r>0$, so we can restrict $r$ to be positive. This set of wave functions constitutes the principal series, and they lead to the continuum that we integrated over in (\ref{primitive}). 

In addition, there could be special values of $\Delta$ with $\text{Re}(\Delta)>\frac{d}{2}$ such that the coefficient of the $|\chi|^{d-\Delta}$ term divided by the coefficient of the $|\chi|^{\Delta}$ term vanishes, leading to a normalizable function. In one dimension this does indeed occur, and the complete set of conformal partial waves includes a discrete set as well as the continuum~\cite{Maldacena:2016hyu}. However, in higher dimensions it does not occur, so the continuum by itself is a complete set. This is established by Theorem 10.5 of~\cite{Dobrev:1977qv} in an abstract way. Here we will show it by analyzing the coefficients explicitly. For simplicity, we assume the spacetime dimension $d$ to be generic. We will be able to describe any integer dimension $d\geq 2$ by taking a limit of generic $d$.

Isolating the factors that can lead to zeros for $\Re(\De)>\frac d 2$, we have
\begin{align}
\frac{K_{\De,J}^{\De_1,\De_2}}{K^{\De_3,\De_4}_{\tl \De,J}} &\propto \frac{1}{\G(d-\De+J)\G(\frac d 2 - \De) (d-\De-1)_J}.
\end{align}
Zeros occur at 
\begin{align}
\De_* = d+J+n, \qquad
\De_* = \frac d 2 + n,
\end{align}
for $n\geq 0$. However, we don't immediately obtain a normalizable state because $G_{d-\De,J}^{\De_i}$ has compensating poles at exactly these locations. Specifically, its pole structure is given by~\cite{Kos:2014bka,Penedones:2015aga}
\begin{align}
G^{\De_i}_{d-\De,J} &\sim -\sum_{n=0}^\oo \frac{c_1(n+1)}{\De-(d+J+n)}G^{\De_i}_{1-J,J+n+1}\nn\\
 &\quad - \sum_{n=1}^\oo \frac{c_2(n)}{\De-\p{\frac d 2 + n}}G^{\De_i}_{\frac d 2+n,J}\nn\\
 &\quad - \sum_{n=1}^J \frac{c_3(n)}{\De-(J+n+1)}G^{\De_i}_{J+d-1,J-n}.
\end{align}
The coefficients $c_1(n+1), c_2(n), c_3(n)$ are given in~\cite{Kos:2014bka}.\footnote{\label{blockfactorfootnote}More precisely, the coefficients in~\cite{Kos:2014bka} are correct for a different normalization of the blocks than we use here, $G_{\Delta,J}^{\text{(here)}} = (-1)^J4^\Delta\frac{\Gamma(\frac{d-2}{2})\Gamma(J+d-2)}{\Gamma(d-2)\Gamma(J+\frac{d-2}{2})}G_{\Delta,J}^{\text{(there)}}$. This implies $c_2(k)^{\text{(here)}}= 4^{-2k}c_2(k)^{\text{(there)}}$ and somewhat more complicated factors of proportionality for $c_1,c_3$.} The poles corresponding to $\De_*=d+J+n$ have residues proportional to the non-normalizable block $G_{1-J,J+n+1}$, so they do not give rise to normalizable states. The poles corresponding to $\De_*=\frac d 2+n$ have normalizable residues $G_{\frac d 2+n,J}$. However, in this case the coefficient function $c_2(n)$ is such that this residue exactly cancels the block $G_{\De,J}^{\De_i}$:
\begin{align}
\lim_{\De\to \frac d 2 + n} \p{G_{\De,J}^{\De_i} + \frac{K^{\De_1,\De_2}_{\De,J}}{K^{\De_3,\De_4}_{d-\De,J}} G_{d-\De,J}^{\De_i}}
&= 0.
\label{eq:absenceofdiscrete}
\end{align}
In other words, our candidate normalizable state vanishes.\footnote{Incidentally, we can turn the logic around: demanding the absence of discrete states (\ref{eq:absenceofdiscrete}) gives a way to determine the coefficient $c_2(n)$, and the cancellation of poles described in the next section gives a method to determine $c_1(n),c_3(n)$. These coefficients are somewhat complicated to compute using other methods~\cite{Kos:2014bka,Penedones:2015aga}.}
This conclusion holds for all $d>1$. When $d=1$, the coefficient $c_1(n+1)$ vanishes, allowing discrete states of the type $\De_*=d+J+n$ to exist.

So we have established that for $d>1$, the principal series wave functions are a complete set of functions symmetric under $\chi\leftrightarrow \b\chi$. The precise completeness relation
\be
\sum_{J=0}^\infty \int_{d/2}^{d/2+i\infty} \frac{d\Delta}{2\pi i} \frac{\Psi^{\De_i}_{\De,J}(\chi,\b\chi)\b\Psi^{\De_i}_{\De,J}(\chi',\b\chi')}{n_{\De,J}} = \frac{\vol(\SO(d{-}2))|\chi|^{2d}}{|\text{Im}(\chi)|^{d-2}}\p{\delta^{(2)}(\chi-\chi') + \delta^{(2)}(\chi-\b\chi')}
\ee
is fixed by taking an inner product with $\Psi^{\De_i}_{\De',J'}$ and using the orthogonality relation (\ref{innerproddef}).

\subsubsection*{Real external dimensions}

We now move to the physically relevant case where the external dimensions are real. There are two approaches we can take. The first approach is to rewrite the completeness relation for the case of principal series external dimensions as
\be
\sum_{J=0}^\infty \int_{d/2}^{d/2+i\infty} \frac{d\Delta}{2\pi i} \frac{\Psi^{\De_i}_{\De,J}(\chi,\b\chi)\Psi^{\tl\De_i}_{\tl\De,J}(\chi',\b\chi')}{n_{\De,J}} = \frac{\vol(\SO(d{-}2))|\chi|^{2d}}{|\text{Im}(\chi)|^{d-2}}\p{\delta^{(2)}(\chi-\chi') + \delta^{(2)}(\chi-\b\chi')},
\ee
where as always $\tl\Delta \equiv d-\Delta$. We can then analytically continue the LHS in $\Delta_i$. In fact, it is convenient to separate the ``block'' and ``shadow'' terms in the first partial wave and then include the shadow term by extending the range of integration over the block term. This leads to an equivalent form
\be
\sum_{J=0}^\infty \int_{d/2-i\infty}^{d/2+i\infty} \frac{d\Delta}{2\pi i} G^{\De_i}_{\De,J}(\chi,\b\chi)\frac{K^{\Delta_3,\Delta_4}_{\tl\De,J}}{n_{\Delta,J}}\Psi^{\tl\De_i}_{\tl\De,J}(\chi',\b\chi') = \frac{\vol(\SO(d{-}2))|\chi|^{2d}}{|\text{Im}(\chi)|^{d-2}}\p{\delta^{(2)}(\chi-\chi') + \delta^{(2)}(\chi-\b\chi')}.
\ee
We can now integrate both sides against the four point function as a function of $\chi',\b\chi'$, including a measure factor $|\text{Im}(\chi')|^{d-2}/|\chi'|^{2d}$. This immediately gives the second line of (\ref{primitive}).

The only subtlety here is that as we continue in the external dimensions, poles in the integrand may cross the contour of integration for $\Delta$. The term that can have poles is the term with the $G_\Delta$ from the remaining partial wave. The coefficient of this term is proportional to $K^{\Delta_3,\Delta_4}_{\tl\De,J}K^{\De_1,\De_2}_{\tl\De,J}$, which includes the factors
\be\label{infactor}
\Gamma\left(\frac{\De+\De_{12}+J}{2}\right)\Gamma\left(\frac{\De-\De_{12}+J}{2}\right)\Gamma\left(\frac{\De+\De_{34}+J}{2}\right)\Gamma\left(\frac{\De-\De_{34}+J}{2}\right).
\ee
When the external dimensions are in the principal series, all poles in this expression are to the left of the contour of integration, but as we continue to real external dimensions with large differences, some poles may cross the line $\Delta = \frac{d}{2}+i\mathbb{R}$. Our analytic continuation prescription instructs us to deform the contour so that the poles do not actually cross it, in other words so that the poles effectively remain to the left of the contour. This has the following important implication. We expect the function $c(J,\Delta) = I_{\Delta,J}K^{\De_3,\De_4}_{\tl\De,J}/n_{\De,J}$ to also inherit the singularities of these gamma functions. When we proceed to deform the contour in (\ref{primitive}) to the right to obtain the OPE, we should not pick up this set of poles.

So far we have discussed the case of real external dimensions by analytically continuing the completeness relation from the case where the external dimensions are in the principal series. An alternative approach is to argue directly for a completeness relation in this case. The first step is to write the bilinear pairing as an inner product, which we can accomplish for real external dimensions by writing
\begin{align}\label{innerprodtwo}
\left(\Psi^{\tl\De_i}_{\tl\De',J'},\Psi^{\De_i}_{\De,J}\right)&=\left\langle \Psi^{\De_i}_{\De',J'},\Psi^{\De_i}_{\De,J}\right\rangle_2,\nn\\
 \langle F,G\rangle_2 &\equiv \frac{1}{2\vol(\SO(d{-}2))}\int \frac{d^2\chi}{|\chi|^{2d}}|\text{Im}(\chi)|^{d-2}|1-\chi|^{-\Delta_{12}+\Delta_{34}}\overline{F}(\chi,\b\chi)G(\chi,\b\chi).
\end{align}
Note the extra factor in the measure, which came from using (\ref{fromdolan}). For small $\Delta_{12},\Delta_{34}$, the Casimir operators are self-adjoint with respect to this inner product. However, for large $\Delta_{12}$ and/or $\Delta_{34}$, the partial waves stop being normalizable, and also the Casimir operators stop being self-adjoint, because of divergences at $\chi =1$ and/or $\chi = \infty$. It is possible that in this case the inner product can simply be modified by defining the integrals by subtracting divergences near $\chi = 1$ and $\chi = \infty$.

In this way of thinking about the completeness relation, the contour prescription described above gets interpreted as including the contribution from a finite number of normalizable discrete states. These discrete states are present if the external dimensions are sufficiently different from each other. They are diagnosed by zeros in the expression
\be
\frac{K_{\De,J}^{\De_1,\De_2}}{K^{\De_3,\De_4}_{\tl \De,J}} \propto \frac{1}{\Gamma(\frac{\De+\De_{12}+J}{2})\Gamma(\frac{\De-\De_{12}+J}{2})\Gamma(\frac{\De+\De_{34}+J}{2})\Gamma(\frac{\De-\De_{34}+J}{2})}.
\ee
for $\text{Re}(\Delta)>\frac{d}{2}$. These are precisely the locations where we encounter poles in the factor (\ref{infactor}). We expect that the contour that avoids the poles as described in the previous treatment can be understood as a contour that includes the principal series and also circles around the discrete states at the locations of the poles.

\section{Subtleties in the Euclidean formula}

\subsection{Spurious poles in the continuation off the principal series}
In order to recover the OPE from the integral over the principal series, one deforms the contour over $\Delta$ in the direction of larger $\text{Re}(\Delta)$. In the process, we pick up the poles representing operators in the OPE\@. However (in addition to the subtlety described in~\ref{onethinghere}), we also pick up two sets of spurious poles: one set from poles in the conformal blocks, and another set from poles in the coefficient function. The fact that these could cancel each other was pointed out in~\cite{Cornalba:2007fs,Costa:2012cb,Caron-Huot:2017vep}. Here we show that the cancellation indeed happens in general, extending an argument from~\cite{Murugan:2017eto}. This may have been implicit in~\cite{Dobrev:1975ru}.

The first set of poles is due to the fact that the conformal block $G^{\Delta_i}_{\Delta,J}$ has a set of poles $\Delta = J+d-1-k$ for $k = 1,\dots,J$, with residues given by $c_3(k)G^{\Delta_i}_{J+d-1,J-k}$, where $c_3(k)$ is defined in~\cite{Kos:2014bka} (up to the convention difference for conformal blocks described in footnote~\ref{blockfactorfootnote}). The contribution to the four-point function from these poles is
\begin{align}
-\sum_{J=1}^\infty\sum_{k = 1}^J&\frac{I_{J+d-1-k,J}}{n_{J+d-1-k,J}}K^{\Delta_3,\Delta_4}_{1+k-J,J}c_3(k)G_{J+d-1,J-k}^{\Delta_i}(x_i) \\&= -\sum_{J=0}^\infty\sum_{k = 1}^\infty\frac{I_{J+d-1,J+k}}{n_{J+d-1,J+k}}K^{\Delta_3,\Delta_4}_{1-J,J+k}c_3(k)G_{J+k+d-1,J}^{\Delta_i}(x_i),\label{firstpoles}
\end{align}
where in the second line we reindexed the summation so that $J$ on the second line is the same as $J-k$ on the first line, note that this substitution should be made for the $J$-dependence in $c_3(k)$ as well.

The second set of poles comes from the factor $\Gamma(d-\Delta+J-1)$ in $K^{\Delta_3,\Delta_4}_{d-\Delta,J}$, which has poles at $\Delta = J+k+d-1$. The pole at $k = 0$ is canceled by a pole in the factor $n_{\Delta,J}$, but for $k = 1,2,\dots,\infty$ we have poles in $K^{\Delta_i}_{\Delta,J}/n_{\Delta,J}$. The residue of  $\Gamma(d-\Delta+J-1) = \Gamma(-k)$ at integer $k$ is $(-1)^{k+1}/\Gamma(k+1)$, and we find the contribution from such poles to the four-point function is
\begin{align}
-\sum_{J = 0}^\infty \sum_{k = 1}^\infty \frac{(-1)^{k+1}}{\Gamma(k+1)}\frac{I_{J+k+d-1,J}}{n_{J+k+d-1,J}}\frac{K^{\Delta_3,\Delta_4}_{1-k-J,J}}{\Gamma(-k)}G^{\Delta_i}_{J+k+d-1,J}(x_i).\label{secondpoles}
\end{align}
Because we have the same set of conformal blocks appearing in (\ref{firstpoles}) and (\ref{secondpoles}), there is the possibility that they cancel. For this to actually happen, we need to find a universal relationship between the theory-dependent factors $I_{J+d-1,J+k}$ and $I_{J+k+d-1,J}$. The necessary relationship follows from an identity between partial waves
\be
\Psi^{\tl\De_i}_{1-J,J+k} = 2^{-k}\frac{\Gamma(J-k+d-2)\Gamma(J+\frac{d-2}{2})}{\Gamma(J+d-2)\Gamma(J-k+\frac{d-2}{2})}\frac{\Gamma(\frac{1-k-\Delta_{12}}{2})\Gamma(\frac{1+k+\Delta_{34}}{2})}{\Gamma(\frac{1+k-\Delta_{12}}{2})\Gamma(\frac{1-k+\Delta_{34}}{2})}\Psi^{\tl\Delta_i}_{1-k-J,J},\label{wavefnrel}
\ee
which holds for $k = 1,2,\dots$. This can be established (working with generic $d$ and external dimensions), using the formulas in~\cite{Kos:2014bka}. The conformal block in the ``shadow'' term in the LHS is proportional to a pole, but the expression is finite because the pole is cancelled by a zero in the coefficient $K$. Similarly, the ``block'' term on the RHS is proportional to a pole that is similarly cancelled. After taking these poles into account, one finds that both of the naively different partial waves actually contain the same two blocks: $G^{\tl\De_i}_{1-J,J+k}$ and $G^{\tl\De_i}_{J+d+k-1,J}$, with specific coefficients so that the above holds.

Now, from the definition (\ref{eucinv}), this relation between the wave functions implies the equation where we replace the wave function on the LHS of (\ref{wavefnrel}) with  $I_{J+d-1,J+k}$ and the one on the RHS with $I_{J+k+d-1,J}$. One can then check that this is precisely what is needed to make sure that (\ref{firstpoles}) and (\ref{secondpoles}) indeed cancel, once we evaluate the other factors of $K$ and $n$ using the explicit formulas in appendix~\ref{app:partialwaves}.

\subsection{Non-normalizable contributions to the four-point function}\label{app:non-norm}
\subsubsection*{Near $\chi = 0$}
The functions $\Psi_{\Delta,J}$ with $\Delta$ in the principal series gives a complete basis of normalizable functions, but the four-point function of a CFT is actually never normalizable in the relevant sense, which requires the function to decay faster than $|\chi|^{d/2}$ for small cross ratios. In particular, the identity operator and scalar operators with $\Delta \le \frac{d}{2}$ (if there are any) give non-normalizable contributions. So, to make sense of the manipulations in this paper, we should subtract these contributions from the four-point function, and then apply the discussion to the normalizable remainder.

A subtlety in this is that to preserve single-valuedness of the four-point function, we need to subtract the full partial wave (block + shadow block) corresponding to the low-dimension scalar operators, not just the conformal block. Since these subtractions involve scalar operators only, they do not spoil the good behavior of the four-point function in the Regge limit. And, in fact, they drop out altogether once we take the double commutator (by our discussion in section~\ref{sec:wickhigherd}). This means that when we use the Lorentzian inversion formula, we do not need to explicitly subtract any contributions for low dimension operators in the $\chi \rightarrow 0$ channel.

To recover the full four-point function, we will have to add back the partial waves that we subtracted, in addition to the integral over the principal series in (\ref{primitive}). The ``block'' parts of these partial waves contain the contributions from physical operators with $\Delta < d/2$ that we expect. However, an apparent puzzle is that they also contain shadow contributions that generically should not be present in the theory. The resolution is that if $I_{\Delta,J}$ is defined with the subtraction procedure described here, then it must contain a pole at the location of the shadow operator. When we shift the contour off the principal series to recover the OPE, we will then get a contribution proportional to the shadow block. This must cancel the explicit shadow part of the partial wave that we add at the end. This can be checked explicitly for the four-point function corresponding to mean field theory.

\subsubsection*{Near $\chi = 1$ or $\chi = \infty$}
In addition, the four-point function may fail to be normalizable near $\chi = \b\chi = 1$ or $\chi = \b\chi = \infty$. For example if all external operators are identical, then from the contribution from the identity operator in the $O_2O_3$ OPE we get a contribution to the stripped four-point function proportional to $|1-\chi|^{-2\Delta_O}$. If $\Delta_O > \frac{d}{2}$ then the four-point function will not be normalizable. In this case, we define $I_{\Delta,J}$ by simply subtracting the divergences, for example by removing a small ball of radius $\epsilon$ around the point $\chi =\b\chi = 1$, doing the integral for fixed $\epsilon$ and subtracting power divergences in $\epsilon$. This will lead to a well-defined expression for $I_{\Delta,J}$. However because the four-point function we are trying to represent is not normalizable, when we try to go back to the four-point function using (\ref{primitive}), the integral over the principal series of $I_{\Delta,J}$ may not converge. The correct prescription is to simply ignore this, and recover the OPE by shifting the contour without worrying about convergence of the principal series integral at $\Delta = \frac{d}{2} \pm i\infty$. This prescription can be justified by showing that it works for the mean field theory correlation function $|\frac{\chi}{1-\chi}|^{2\Delta_O}$ and then subtracting and adding the mean field theory answer to the physical four-point function.

This situation is very analogous to the fourier transform $\int dx e^{ipx}|x|^{-a}$ for $a>1$. We can define the integral by analytic continuation in $a$ or equivalently by removing an interval of size $\epsilon$ around the origin and subtracting divergences. We find a multiple of $|p|^{a-1}$. When we try to reverse this and compute $\int dp e^{-ipx} |p|^{a-1}$, the integral is not convergent for large $p$, but we get the right answer by nevertheless shifting the contour either into the right or left half-plane of $p$, depending on the sign of $x$, and doing a convergent integral along the branch cut.

\section{A different way of obtaining $B_J(\eta)$}\label{app:differentB}
In this appendix we will describe a second way of passing from the Gegenbauer polynomial $C_J(\eta)$ to the better-behaved function $B_J(\eta)$. In other words, we will give an alternate route from (\ref{eq:kernelformulaforc}) to (\ref{averaged}). The method discussed in this section is more closely related to the usual treatment of the functions $C_J(\eta)$ and $B_J(\eta)$ when studying amplitudes, see e.g.~\cite{Paulos:2017fhb}.

We begin by further gauge-fixing the expression for $I_{\Delta,J}$ in (\ref{eq:kernelformulaforc}) by setting
\be\label{gauge2}
x_3 = (x,\tau,y,0,\dots), \hspace{20pt} x_4 = (x',\tau',y,0,\dots).
\ee
Note that we take the third coordinates of $x_3$ and $x_4$ to be equal. The Faddeev-Popov determinant for this gauge-fixing is proportional to $\frac{1}{4}|y|^{d-3}|\tau - \tau'|^{d-2}$, so we obtain
\be
I_{\Delta,J} = \int_{-\infty}^\infty \frac{dx dx' d\tau d\tau' dy }{4\vol(\SO(d{-}3))}\frac{|y|^{d-3}|\tau-\tau'|^{d-2}}{|x_{34}|^{2d-\Delta_3-\Delta_4-\Delta}}\langle O_1O_2O_3O_4\rangle\hat C_J\left(\eta\right),
\ee
where $\eta$ reduces in this gauge to $\eta = (x-x')/|x_{34}|$.
At this point, we would like to Wick-rotate in $\tau,\tau'$. However, if we try to do this with the integrand in its current form, we will have two problems. One problem is that the argument of the Gegenbauer polynomial will become large in places, and $\hat{C}_J$ grows for large arguments. The second problem is that in odd dimensions the integrand isn't analytic to begin with, because of the factor $|\tau-\tau'|^{d-2}$.

We can avoid both of these problems as follows. Let's begin by considering the function $B_J(\eta)$, defined in (\ref{eq:defofBJ}). This is a solution to the same Gegenbauer differential equation as $\hat C_{J}$, but it is not a polynomial. {Instead, as is clear from (\ref{eq:defofBJ}), it has a branch cut running between $\eta = \pm 1$. A useful fact is that when we consider $-1<\eta<1$, we have
\be\label{CfromB}
\hat C_J(\eta) = i^{d-2} \frac{\hat C(1)}{\vol(S^{d-2})} \left[B_J(\eta+i\epsilon) + (-1)^d B_J(\eta-i\epsilon)\right].
\ee
When $d=3$, this is equivalent to the well-known relationship between Legendre $P$ and $Q$ functions. In general, it can be derived from the integral representations (\ref{eq:gegenbauerrep}), (\ref{eq:bcalculation}):
\begin{align}
\hat{C}_J(\eta) &= \frac{\hat C_J(1)}{\vol(S^{d-2})}\vol(S^{d-3})\int_0^\pi d\theta\sin^{d-3}\theta \frac{(x+i\tau\cos\theta)^J}{(x^2+\tau^2)^{J/2}}, \hspace{20pt} \eta = \frac{x}{(x^2+\tau^2)^{1/2}},\label{Cintrep}\\
B_J(\eta) &= \vol(S^{d-3})\int_0^{\text{arccosh}(x/t)}d\beta \sinh^{d-3}\beta \frac{(x-t\cosh\beta)^J}{(x^2-t^2)^{J/2}},\hspace{20pt} \eta = \frac{x}{(x^2-t^2)^{1/2}}.
\end{align}
To get an integral expression for $B_J(\eta+i\epsilon)$ with $-1<\eta<1$, we rotate $t$ in the lower half-plane to $t \rightarrow -i\tau$. To get $B_J(\eta-i\epsilon)$, we rotate in the upper half-plane to $t \rightarrow i\tau$. Now, in order to show (\ref{CfromB}), the idea is to break up the integral over $\theta$ in (\ref{Cintrep}) as
\be
\int_0^\pi d\theta = \int_0^{\frac{\pi}{2}-i\text{arcsinh}(x/\tau)}d\theta + \int_{\frac{\pi}{2}-i\text{arcsinh}(x/\tau)}^\pi d\theta.
\ee
In the first term we make the change of variables $\theta = i\beta$, and in the second term we make the change of variables $\theta = \pi + i\beta$. These terms then become exactly the integral representations of the two $B_J$ functions on the RHS of (\ref{CfromB}), with the continuations just described.}
\begin{figure}
\begin{center}
\includegraphics[width=\textwidth]{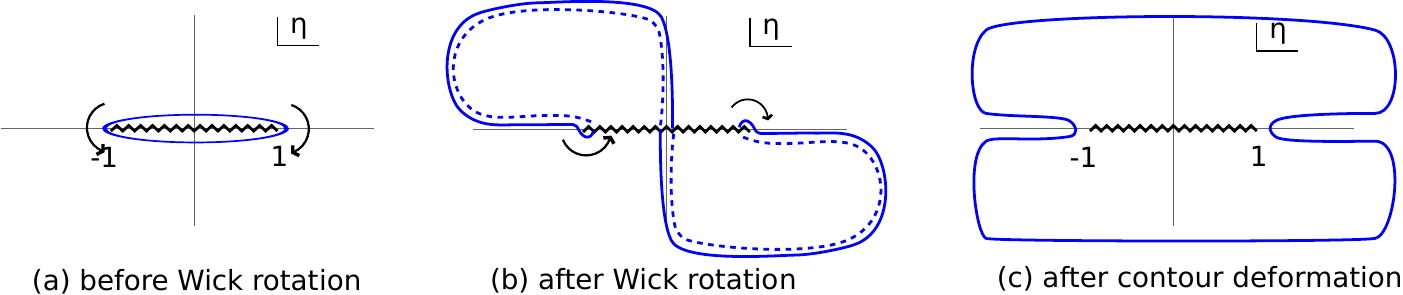}
\caption{\small{{(a) after the replacement (\ref{rep1}), the contour for $\eta$ circles the branch cut of the $B_J$ function. The arrows indicate the direction in which the contour passes the branch points at $\eta = \pm 1$ as we increase $\tau - \tau'$. After Wick rotation we end up deforming the contour as in (b). The dashed parts of the contour are on the second sheet. Note that the arcs at infinity are shown at finite radius for clarity. The arrows indicate the direction in which the contour passes the branch points as we increase $v-v'$, with $u-u'$ fixed. Finally, when we deform the contour over $v,v'$ we pull the dashed portions of the contour on the second sheet back through the cut to the first sheet, giving the contour in (c). We further drop the arcs at infinity, so that we have just the integrals along the real axes, picking up discontinuities across branch cuts from the four point function.}}}\label{fig:contoursforeta}
\end{center}
\end{figure}

Now, in the integral (\ref{gauge2}), the argument of the $\hat C_J$ function ranges between minus one and one. If we change the argument slightly, so that
\be\label{rep1}
\eta  = \frac{x-x'}{\sqrt{(x{-}x')^2+(\tau{-}\tau')^2}} \rightarrow \frac{x-x' + i \epsilon\, \text{sgn}(\tau'-\tau)}{\sqrt{(x{-}x')^2+(\tau{-}\tau')^2}} 
\ee
then the argument will circle around the interval $-1<\eta<1$. For positive $\tau'-\tau$ we will be above the cut, and for negative $\tau'-\tau$ we will be below it. This means that if we make the replacement (\ref{rep1}) we can also replace $\hat C_J(\eta) \rightarrow [2 i^{d-2}\hat C_J(1)/\vol(S^{d-2})] B_J(\eta)\text{sgn}(\tau'-\tau)^d$ and we will get the same answer. The nonanalytic factor of the $\text{sgn}$ function is needed because of the $(-1)^d$ in (\ref{CfromB}). Happily, this factor combines with a factor from the measure to give an analytic integrand. In other words, we have justified the replacement
\be
|\tau-\tau'|^{d-2}\hat C_J(\eta) \rightarrow \frac{2 i^{d-2}\hat{C}_J(1)}{\vol(S^{d-2})}(\tau'-\tau)^{d-2}B_J(\eta),
\ee
where $\eta$ is understood with the $i\epsilon$ prescription in (\ref{rep1}). Note that at this point we are still in Euclidean signature, the $i\epsilon$ is simply to guide our integral around the branch cut of the $B_J$ function.

We now have an analytic integrand, and the $B_J$ function is decaying at large argument, so at this point we can Wick-rotate in $\tau,\tau'$. Most of this follows closely the discussion in the main text of the paper. However, there is one potential subtlety. After Wick-rotation, we would like to deform the $v,v'$ contours in either the upper or lower half-planes as in figure~\ref{fig:contourDef} to get the double-commutator expression. The $B_J$ function has branch points at $\eta = \pm 1$, and a possible concern is that these singularities might lie in the half-plane we are trying to deform through. In fact, this does not happen, the branch point singularity is always in the half-plane that we are not deforming in. {This is explained by the arrows in figure~\ref{fig:contoursforeta}, which show the direction the contour passes around the branch points. Let's consider the case $u - u'>0$. Then as we vary $v-v'$ we approach the branch point at $\eta = 1$, but we pass around it in a clockwise manner as $v-v'$ increases. This implies that the singularity is in the lower half-plane for $v-v'$, and we are free to deform this variable in the upper half-plane, as we did in the main text. The argument when $u-u'$ is negative is similar; the contour passes by the branch point at $\eta = -1$ in a counterclockwise manner, which means the singularity is in the upper half-plane for $v-v'$.}

Finally, after the Wick rotation and contour deformation, there are two regions that we integrate over, as in figure~\ref{fig:regionsR1R2} (but with curved boundaries for nonzero transverse separation $y$). In the region where we deform the contours to get $\langle [O_4,O_2][O_3,O_1]\rangle$, we have $t-t'>0$ and $\eta >0$, and so $i^{d-2}(\tau'-\tau)^{d-2} = |t-t'|^{d-2}$. In the region where we close the contours to get $\langle [O_3,O_2][O_4,O_1]\rangle$, we have $t-t' <0$, and $\eta <0$, so $i^{d-2}(\tau'-\tau)^{d-2}= (-1)^d |t-t'|^{d-2}$. Using $B_J(-x) = (-1)^{d+J}B_J(x)$ we can write
\be
i^{d-2}(\tau'-\tau)^{d-2}B_J(\eta) = (-1)^J|t-t'|^{d-2}B_J(-\eta).
\ee
The expression in these two regions can now be recognized as gauge-fixed versions of the two terms in (\ref{averaged}), for which the determinant is proportional to $\frac{1}{2}|y|^{d-3}|t-t'|^{d-2}$. We can therefore proceed from that point in the main derivation, having skipped there from (\ref{eq:kernelformulaforc}).

\section{Subtleties in the Lorentzian formula}
\subsection{No extra singularities during the $v$ contour deformation}
In sections~\ref{sec:contourmanipulation} and~\ref{sec:wickhigherd} we used a contour deformation in the null coordinates $v_3,v_4$ to go from an integral over all of Lorentzian space to an integral of the double commutator over a region defined by lightcones. This argument would be spoiled if we encounter singularities in the four-point function as we make this contour deformation (other than the singularities at null separation of external points that give the double commutator itself). In general, four-point functions can indeed have additional singularities. These come from Landau-like diagrams~\cite{Maldacena:2015iua}, somewhat similar to the Landau diagrams that generate singularities in scattering amplitudes~\cite{Landau:1959fi,Coleman:1965xm,Cutkosky:1960sp}. In this section we will argue that our contour manipulations are still safe.

The argument is as follows. For complex values of $v$, we can formally write
\be\label{notwell}
O(v) = O(v_R+i v_I) = e^{v_I P_v}O(v_R) e^{-v_I P_v},
\ee
where $P_v\le 0$ is the non-positive operator generating translations in the $v$ direction. In general, (\ref{notwell}) is not well defined, since one or the other of the exponential factors will be unbounded. However, if $v_I >0$ so that we are in the upper half-plane, then (\ref{notwell}) makes sense acting on the vacuum, $O(v)|0\rangle$ because $e^{-v_I P_v}$ gives one acting on the vacuum, and the $e^{v_IP_v}$ operator is bounded for $v_I\ge 0$. Also, in vacuum correlation functions in which $O(v)$ is ordered first (rightmost) in the list of operators, we can give the operator an $i\epsilon$ prescription with respect to a timelike direction, further replacing $O(v) \rightarrow e^{-\epsilon H}O(v)e^{\epsilon H}$. After doing this, one can show that the correlation function will be analytic in the upper half plane for $v$. Similarly, correlation functions in which $O(v)$ is ordered last (leftmost) will be analytic in the lower half-plane.

Now one simply has to check that for the continuations used in sections~\ref{sec:contourmanipulation} and~\ref{sec:wickhigherd}, the correlation function can be written with an operator ordering consistent with the half-plane in which we deform $v$. Which half-plane we want to use for e.g.~the $v_4$ coordinate depended on the relative ordering of the $u$ coordinates. Up to discrete symmetries, there are two cases to consider. First, suppose that $u_4$ is the largest of the $u$ coordinates. Then $x_4$ is either spacelike or in the past of the other points, so we can write the correlation function with $O_4$ ordered first, next to the vacuum, and we have analyticity in the upper half $v_4$ plane. And, indeed, in this situation our argument required us to deform $v_4$ in the upper half plane in order to get zero for the integral. The other case to consider is when $u_4 < u_1$ but $u_4$ is larger than $u_2,u_3$. Then in the region of the $v_4$ integral such that $x_4$ is spacelike or to the past of $x_1$, we can again write the correlator with $O_4$ ordered first, and deform in the upper half plane, but in the region of the $v_4$ integral where $x_4$ is in the future of $x_1$ we cannot. This precisely allows the contour deformation that we followed (see the right panel of figure~\ref{fig:contourDef}), where we leave the contour where it is for $v_4$ large enough that $x_4$ is in the future of $x_1$, but we deform the contour in the upper half plane for smaller values of $v_4$.

\subsection{The requirement that $J>1$}\label{app:reqJgtr1}
We should also check that the contributions near infinity can be dropped after doing the Wick rotation. The discussion in section~\ref{sec:wickhigherd} is less convenient for addressing this question, since the integral after Wick rotation is only conditionally convergent: we have to do the $v$ integral before the $u$ integral. Instead we will use the perspective in appendix~\ref{app:differentB} where we pass to the $B_J(\eta)$ function before Wick rotation. The important question is whether we can drop the parts of the integral corresponding to the arcs near infinity in figure~\ref{fig:contoursforeta}. These regions correspond to parts of the integral where $x_3$ and $x_4$ are almost null separated from each other and $\eta$ diverges. The thing in our favor is that for large spin, $B_J(\eta)$ is a rapidly decreasing at large $\eta$, providing convergence. The question is how large is large enough? We expect to find at most a power law singularity in the integrand, which has equal strength when approaching from any direction. This means that we can drop the arcs at infinity if the integral along the real axis (i.e.\ we keep after dropping the arcs) is convergent.

What this means is that our manipulations are justified if our final formula is convergent separately for each of the terms that appear in the double commutator. We should check that this is the case assuming the dimension $\Delta$ is in the principal series $\Delta = \frac{d}{2} + ir$, so that the original Euclidean inversion formula (\ref{eucinv}) makes sense. It is convenient to assess the convergence using the formula expressed in terms of cross ratios, e.g.~(\ref{finalI}). The dangerous region (corresponding to $x_3$ and $x_4$ almost null separated) is small $\chi,\b\chi$. If we take both to zero simultaneously, $\b\chi \sim \chi$, then $H_{\Delta,J}(\chi,\b\chi)$ is proportional to $\chi^{J+d-1}$. The correlation function (for any of the orderings) divided by $T^{\Delta_i}$ is bounded by a constant in this limit~\cite{Maldacena:2015waa,Hartman:2015lfa}, so from the behavior of the measure we conclude that to have convergence we need $J > 1$. Another limit to consider is small $\chi$ with $\b \chi$ fixed. In this light-cone limit, after subtracting the contribution from the identity, the correlation vanishes as $\chi^{\tau/2}$ where $\tau \ge \frac{d-2}{2}$ is the smallest twist of the theory. Combining with the measure and the block $H_{\Delta,J}$ for $\Delta$ in the principal series, we again find that $J>1$ is sufficient.

What this means is that the formula (\ref{finalI}) gives the same answer as the Euclidean formula for $I_{\Delta,J}$ for all spins $J = 2,3,\dots$. However, for $J = 0,1$ the formula could diverge or give an answer that differs from the correct Euclidean expression. {A small subtlety here is that the above statements may not commute with the $1/N$ expansion. In the Regge limit, the $1/N^2$ term in the four point function can grow. The chaos bound implies that the above manipulations would still be valid for $J>2$. At higher orders in the $1/N^2$ expansion we expect further restrictions on $J$. However, if we study the exact finite $N$ correlator rather than its $1/N^2$ expansion, the only requirement is that $J>1$.}

{We can understand the fact that the formula only applies to $J>1$ in another way. One can add partial waves to the four point function with $J = 0,1$ without spoiling boundedness in the Regge limit. However, the double commutator of such partial waves vanishes, so they will make no contribution to our Lorentzian formula for $I_{\Delta,J}$. This means that this formula does not in general correctly capture the contributions with $J = 0,1$.}

Note that as we continue $\Delta$ off the principal series the integral will not in general be convergent. This doesn't indicate a failure of our continuation argument, it only means that the continuation of $I_{\Delta,J}$ in $\Delta$ has poles. These poles represent the physical operators of the theory, as described in the Introduction.

\bibliography{references}

\bibliographystyle{utphysmodb}

\end{document}